\documentclass[%
 aip,
 amsmath,amssymb,
 preprint,%
 superscriptaddress,
]{revtex4-2}

\usepackage{graphicx}
\usepackage{float}
\usepackage{dcolumn}
\usepackage{bm}

\usepackage[normalem]{ulem} 
\usepackage{soul} 
\usepackage{xcolor}
\sethlcolor{yellow}

\usepackage[utf8]{inputenc}
\usepackage[T1]{fontenc}
\usepackage{mathptmx}
\usepackage{etoolbox}

\newcommand{\sca}[1]{\textit{#1}}
\newcommand{\vect}[1]{\textbf{\fontfamily{ptm}\selectfont #1}}
\newcommand{\matr}[1]{\textbf{\fontfamily{phv}\selectfont #1}} 
\newcommand{\unitvector}[1]{\hat{#1}}

\makeatletter
\def\@email#1#2{%
 \endgroup
 \patchcmd{\titleblock@produce}
  {\frontmatter@RRAPformat}
  {\frontmatter@RRAPformat{\produce@RRAP{*#1\href{mailto:#2}{#2}}}\frontmatter@RRAPformat}
  {}{}
}%
\makeatother
\begin{document}

\preprint{AIP/123-QED}

\title{Exceptionally Large Fluctuations in Orientational Order:\\The Lessons of Large-Deviation Theory for Liquid Crystalline Systems}

\author{Eleftherios Mainas} 
\author{Richard M. Stratt}
    \affiliation{Department of Chemistry, Brown University, Providence, Rhode Island 02912, USA}
\email{richard\_stratt@brown.edu}

\date{\today}

\begin{abstract}

How condensed-matter simulations depend on the number of molecules being simulated ($N$) is sometimes itself a valuable piece of information. Liquid crystals provide a case in point. Light scattering and $2d$-IR experiments on isotropic-phase samples display increasingly large orientational fluctuations (“pseudo-nematic domains”) as the samples approach their nematic phase. The growing length scale of those locally ordered domains is readily seen in simulation as an ever-slower convergence of the distribution of orientational order parameters with $N$. But the rare-event character and exceptionally slow time scales of the largest fluctuations make them difficult to sample accurately. We show in this paper how taking a large-deviation-theory perspective enables us to leverage simulation-derived information more effectively. A key insight of the theory is that finding quantities such as orientational order parameters (extensive variables), is completely equivalent to deducing the conjugate (intensive) thermodynamic field required to equilibrate that amount of order - and that knowing the relationship between the two (the “equation of state”) can easily be turned into knowing the relative free energy of that degree of order. A variety of well-known thermodynamic integration strategies are already founded on this idea, but instead of applying an artificially imposed external field, we use a priori statistical mechanical insights into the small and large-field limits to construct a simulation-guided, interpolated, equation of state. The free energies that result mostly need information from the most probable configurations, making the simulation process far more efficient than waiting for (or artificially generating) large fluctuations.

\end{abstract}

\maketitle

\section{\label{sec:level1}INTRODUCTION}

It has long been appreciated that isotropic liquid crystalline systems exhibit increasingly large fluctuations in nematic order when the systems approach their isotropic/nematic (I/N) phase transitions. \cite{degennes, Gennes} Early light scattering experiments, \cite{stinson1970pretransitional, litster1970critical} modern video microscopy experiments \cite{zhang2011, mishra2014} 
as well as a plethora of 2\textit{d}-IR experiments \cite{sokolowsky2013dynamics, sokolowsky2014length, sokolowsky2014new, sokolowsky2016critical, cang2002short, stankus1992nanosecond, li2006mode, li2006three} have measured the orientational relaxation dynamics of isotropic systems close to the isotropic/nematic transition. The key observation in all these studies was the rapid growth of correlation times and lengths as pseudo-nematic domains start to form inside the isotropic phase. Traditional Landau-de Gennes theories ascribed this phenomenon to increasing proximity to a hidden critical point, \cite{Gennes, degennes, chakrabarti2009, gramsbergen} albeit one that is never actually reached because it is preempted by the first-order I/N phase transition.  Simulations show this same phenomenon as a finite size effect: \cite{greschek2011, priezjev2001, weber1999, zhao} an increasingly slow convergence to the expected Gaussian portrait of the fluctuation distribution with increasing simulation size.

Furthermore there are distinct characteristics of the two-dimensional case that need to enter the analysis. Because of the Mermin-Wagner theorem, \cite{mermin} we might expect thermal fluctuations to destabilize both translational and orientational long-range order in two dimensions. That would mean that for large enough systems, local patches of liquid
orientational (nematic) order could persist without ever achieving a system-wide consensus as to the orientation of the 
director. The orientationally ordered liquid
in 2d might therefore more accurately called quasi-nematic.\cite{Eppenga1985} A number of analytical and simulation approaches\cite{Eppenga1985, ref-y, ref-z, vink2009isotropic, vink2014} have shown that that is, in fact, the case for
intermolecular potentials allowing for realistic coupling between reorientation and
translation. However, the Mermin-Wagner argument is subtle in this situation, with the nature of the phase transition sensitive to the form and level of anisotropy of the intermolecular interactions.\cite{ref-x, ref-y, ref-z} Moreover, as in two-dimensional systems lacking the possibility of orientational order, a Kosterlitz-Thouless phase transition\cite{Kosterlitz1973, ref-aa} could intervene between the isotropic and quasi-nematic regimes.\cite{ref-z} Yet additional complications arise from the fact that ellipsoidal molecules in two dimensions are notorious glass formers. In fact, a two-step glass transition, one in which the translational and orientational degrees of freedom freeze out separately, has been observed experimentally via video microscopy techniques.\cite{glass2d-paper1, glass2d-paper2} Recent experiments, though,\cite{glass2d-paper2, SeanLing2021} have managed to bypass the glassy metastabilities and have confirmed the existence of the pseudo-nematic ordering predicted by numerical simulations.\cite{frenkel, Eppenga1985, ref-y, ref-z} Two-dimensional studies with Gay-Berne interactions having an aspect ratio of 3 (the same, more-or-less realistic, potential model we apply our methods to in this paper)\cite{Kundu, demiguel} have found a transition between an isotropic phase (where orientational correlations decay exponentially) and a pseudo-nematic phase (in which those same correlations have a power law decay).\cite{Kundu, KUNDU2021113224} But the character of the incipient nematic fluctuations in a
macroscopically isotropic liquid remains a topic of continuing interest.

We want to understand the growth of such nematically ordered domains in putatively isotropic liquid crystals both in two and three dimensions. In a simulation context, though, such fluctuations are always going to be difficult-to-sample rare events. Can simulations nonetheless be used to explore these large fluctuations?

We suggest here that the machinery of equilibrium \textit{large-deviation theory} \cite{touchette, touchette2012basic} provides a useful way to not only think about the nature of such infrequent events but to efficiently characterize them using the ingredients provided by numerical simulation.  The key insight is that there are two regimes in these problems that are difficult but provide entirely opposite kinds of challenges:  one is the \textit{small-fluctuation} regime.  Intermolecular correlations may be inescapably important in this regime, but the statistics are still governed by the central-limit theorem, so probability distributions are fundamentally Gaussian.  The extreme \textit{large-fluctuation} limit, by contrast, is explicitly non-Gaussian but it is possible to show that it is dominated by single-molecule events, meaning that intermolecular correlations are only of secondary concern.

What large-deviation theory does is give us a way of viewing the interesting large (but not extreme) fluctuation behavior as a well-defined interpolation between these two relatively simple limits – and of constructing that interpolation from information relatively easy to sample from numerical simulations.  In particular, we will find that we do not need to simulate large fluctuation events in our liquid crystals to gain a quantitative understanding of their probabilities.

\section{\label{sec:level1}LARGE DEVIATIONS THEORY AND THE THERMODYNAMIC VIEW OF THE PROBABILITY DISTRIBUTION OF ORDER PARAMETERS}

The formulation of large-deviations theory used to attack non-equilibrium problems (the Freidlin-Wentzell formalism)\cite{touchette, touchette2012basic} has been a frequent recent source of insight in portraits of the glass transition as a dynamical transition. \cite{jack2011, jack2013large, nemoto2017finite, chandler} But it is the more prosaic, equilibrium, counterpart of that theory (the Gartner-Ellis formalism) \cite{touchette, touchette2012basic} that this paper relies on to study improbably large fluctuations. This equilibrium version has recently proven useful in deriving fluctuation relations for systems with broken symmetries.\cite{Gaspard_2012, lacoste_2014, lacoste2015fluctuation, Guioth_2016}

The central point that makes equilibrium large deviation theory applicable to probability distributions of quantities such as orientational order parameters is that these quantities are extensive (proportional to the number of molecules, \sca{N}, when \sca{N} is large).  That simple fact means that there is a very thermodynamic-looking structure to the problem of finding the probability distribution of order parameters.  We briefly sketch that structure here. 

To begin with, finding the probability density of some order parameter value \sca{M} can be thought of as finding a potential of mean force (a free energy per particle) associated with that value, \sca{I}.  In large deviation theory, that potential of mean force is commonly called the rate function.  Large deviation theory then notes that a complementary way to view this potential of mean force is as a function of an intensive “field” variable that biases the possible value of the extensive order parameter, rather than a function of the order parameter itself.  In the large-\sca{N} limit, this bias turns into a unique relationship between the optimum field values and the order-parameter values, what in thermodynamics one would call an equation of state. That in turn can be shown to make the rate function simply the reversible work \cite{callen} required to achieve a given value of the order parameter in the face of these particular fields.

This same kind of Legendre transform has actually been used to good effect in computing how free energy depends on both macroscopically defined states (Landau free energies)\cite{Ilg2011, Ilg2012, Gupta2013, Luo2014, semenov2024} and microscopically defined states \cite{thermoint1, thermoint2, thermoint3, thermoint4, tuckerman2011} (potentials of mean force). But it is worth remembering that such thermodynamic equivalences are guaranteed only at the thermodynamic limit, not in finite-sized simulations. What if one wants to study finite-size scaling, as we do here?\cite{semenov2024} What large deviation theory does is show how these ensemble relations continue to work at any (large) finite $N$, making clear precisely how the $N$ dependence enters. The generality of the large-deviation connections between different finite-$N$ ensembles appears not to have been fully appreciated. It is well known,
for example, how large deviation theory motivates the introduction of an intensive \textit{heat} variable (temperature) in transforming from finite-$N$ microcanonical to canonical free energies, \cite{touchette} but what we do here (and what is done in the literature without any mention of large deviation theory)\cite{Ilg2011, Ilg2012, Gupta2013, Luo2014} is essentially introducing an intensive \textit{work} variable in order to pivot between finite-$N$ Helmholtz and Gibbs free energies.

The basic large deviation theory result we use is the Gärtner-Ellis theorem. \cite{touchette, touchette2012basic}  In a statistical mechanics context, we can write this result in the following way:  If we have a large number of degrees of freedom, $\left(j = 1,...,\sca{N} \right)$, the probability density $\sca{P(M)}$ of an extensive quantity $\sca{M}$ is governed by a rate function $\sca{I(m)}$ 
\begin{equation}
    \sca{M} \equiv \sca{Nm}, \quad \sca{P(M)} \sim \exp{\left(-\sca{N I(m)} \right)} \tag{2.1}\label{eq:2.1}
\end{equation}
which, in turn, can be expressed in terms of an intensive conjugate field $\sca{h}$ and the associated cumulant generating function $\sca{$\lambda$(h)}$
\begin{equation}
    \sca{I(m)} = \sca{mh*} - \sca{$\lambda$(h*)}, \quad \exp{\left(\sca{N} \sca{$\lambda$(h)} \right) = \langle \exp{\left(\sca{hM}\right)} \rangle} \tag{2.2}\label{eq:2.2}
\end{equation}
provided the field is chosen to be take on the value \sca{h*} that maximizes the rate function, namely the value that satisfies
\begin{equation}
    \sca{m} = \left( \frac{\partial \sca{$\lambda$}}{\partial \sca{h}}\right)_{\sca{h=h*}} \tag{2.3}\label{eq:2.3}
\end{equation}
The existence of thermodynamic analogies to these last equations is hard to miss, but comparatively little use has been made of the idea that this last equation is really an equation of state.  From that perspective, the rate function $\sca{I}$ can be thought of as a kind of Helmholtz free energy per particle (a natural function of the extensive work variable $\sca{M}$), and the cumulant generating function $\sca{$\lambda$}$ regarded as a kind of Gibbs free energy per particle (a natural function of the conjugate intensive work variable $\sca{h}$) with the two related by a Legendre transform
\begin{equation}
\sca{d$\lambda$} = \sca{m dh} \Rightarrow \sca{dI = d(mh - $\lambda$) = h dm} \tag{2.4}\label{eq:2.4}
\end{equation}
so that rate function can always be written (to within an additive constant) as the corresponding reversible work
\begin{equation}
    \sca{I} = \int_{0}^{\sca{m}} \sca{h}(\sca{m}')\sca{dm}' \tag{2.5}\label{eq:2.5}
\end{equation}

To apply these ideas to liquid crystals, we will actually make use of the fact that much the same structure, and the same kind of answers, appear when the order parameter is a vector quantity. \cite{touchette}
\begin{equation}
    \vect{M} \equiv \sca{N} \vect{m}, \quad \sca{P}(\vect{M}) \sim \exp{\left(- \sca{N} \sca{I}(\vect{m}) \right)} \tag{2.6}\label{eq:2.6}
\end{equation}
\begin{equation}
    \sca{I}(\vect{m}) = \vect{m}\cdot\vect{h*} - \sca{$\lambda$}(\vect{h*}), \quad \exp{\left(\sca{N} \sca{$\lambda$}(\vect{h}) \right) = \langle \exp{\left(\vect{M}\cdot\vect{h}\right)} \rangle} \tag{2.7}\label{eq:2.7}
\end{equation}
\begin{equation}
    \vect{m} = \left[ \nabla_{\vect{h}} \sca{$\lambda$} (\vect{h}) \right]_{\vect{h}=\vect{h*}}, \quad \sca{I} = \vect{$\int$}_{\vect{0} \rightarrow \vect{m}} \vect{h}(\vect{m}') \cdot d\vect{m}' \tag{2.8}\label{eq:2.8}
\end{equation}

\textit{How this differs from conventional thermodynamic integration strategies}: This formal mathematical framework immediately brings to mind a standard computational approach to studying rare, large-fluctuation, events: instead of waiting for such fluctuations to appear spontaneously in a simulation, the fluctuations can be driven to appear by stepwise applications of progressively larger values of some suitably chosen artificial field.  The potential of mean force can then be computed by adding up the work associated with each step, what is commonly called a “thermodynamic integration” strategy (Thermodynamic integration is routinely used to calculate free energies in the context of liquid crystals, \cite{Ilg2011, Ilg2012, Gupta2013, Luo2014}, solid-state materials, \cite{thermoint1} polymer elasticity, \cite{thermoint4, thermoint2} and a host of chemical and biological processes. \cite{thermoint3, tuckerman2011})
In many of these applications, the field being used does not have to correspond to any physically attainable force and the quantity whose probability is desired does not have to be extensive in order to make the calculation valid; the procedure just parameterizes an arbitrarily chosen, but computationally convenient, transformation between the desired starting and ending thermodynamic states.  One is simply taking advantage of the mathematics allowed by such transformations to compute changes in free energy.\cite{tuckerman2011}

Our approach achieves its goal via a thermodynamic integration as well, but it does so via a field that has a physical significance that we use to our advantage.  Traditional thermodynamic integration calculations compel us to carry out a series of equilibrated simulations over the entire range of fluctuation sizes.  Such simulations are eminently feasible for local rearrangements, but can be problematic when the large fluctuations envisioned are system-wide in scope.  In particular, consider what would happen if we tried to study the orientational ordering of a liquid crystalline system in this way.  Trying to simulate a field large enough to achieve a significant degree of orientational order without simultaneously inducing unwanted translational order might not even be possible; one could easily end up turning the system into a solid.

What we shall do is to find a way to extract information from simulations carried out at just the thermodynamic states of interest and not under special conditions designed to artificially stabilize large fluctuations.  We show how our perspective on large deviation theory provides a route to do just that.

\section{\label{sec:level1}THE STRONGLY-CORRELATED/WEAK-FLUCTUATION AND WEAKLY-CORRELATED/STRONG-FLUCTUATION LIMITS}

Equation (2.2) implies that in order to be able to make any practical use of large deviation theory, we need to be able to first evaluate the cumulant generating function $\sca{$\lambda$}$ as a function of the field $\sca{h}$ -- and for strongly correlated systems, the general problem of evaluating the cumulant generating function, may be no easier than evaluating the desired probability distribution itself.  However, we note that there is quite a bit known about the behavior of the cumulant generating function for both small and large values of the special field that enters Eq. (2.2).

When the field is small enough, the generating function can always be expanded in cumulants (as long as we are not in the immediate vicinity of a critical point),
\begin{equation*}
    \sca{N} \sca{$\lambda$}_{\text{weak-field}}(\sca{h}) = \sca{h} \langle \sca{M} \rangle + \frac{1}{2} \sca{h}^2 \langle \left( \sca{$\delta$} \sca{M} \right)^2\rangle 
\end{equation*}
(where the notation $\sca{$\delta$} \sca{x} = \sca{x} - \langle \sca{x} \rangle$ refers to a fluctuation) or, equivalently
\begin{equation}
    \sca{N} \sca{$\lambda$}_{\text{weak-field}}(\sca{h}) = \sca{h} \langle \sca{M} \rangle + \frac{1}{2} \sca{h}^2 \sca{$\chi$}, \quad \sca{$\chi$} = \langle \left( \sca{$\delta$} \sca{M} \right)^2\rangle \tag{3.1}\label{eq:3.1}
\end{equation}
with $\sca{$\chi$}$ the susceptibility.  It is easy to see from Eqs. (2.2) and (2.3) (and well known from basic probability theory) that this limit corresponds to Gaussian (central-limit-theorem) behavior for the desired probability distribution
\begin{equation}
    \sca{h*} = \frac{\left(\sca{m} - \langle \sca{m} \rangle\right)}{\sca{$\chi$}} \Rightarrow \sca{I}_{\text{weak-field}}(\sca{m}) = \frac{1}{2} \frac{\left(\sca{m} - \langle \sca{m} \rangle\right)^2}{\sca{$\chi$}} \tag{3.2}\label{eq:3.2}
\end{equation}
The development makes clear that this limit paints an intrinsically small-fluctuation picture, but it imposes no restrictions on how strongly correlated the system is.  The susceptibility $\sca{$\chi$}$ could, for example, reflect the effects of significant intermolecular correlations.

Consider, by contrast, the large-field limit, and suppose that the extensive quantity $\sca{M}$ being targeted is a sum of identical (but not necessarily independent) single-molecule contributions, $\sca{x}_j$ , $(j = 1, ..., \sca{N})$
\begin{equation}
    \sca{M} = \sum_{j=1}^{\sca{N}} \sca{x}_j \tag{3.3}\label{eq:3.3}
\end{equation}
The crucial observation is that a strong enough field will always dominate the intermolecular correlations, regardless of how strong the correlations are:  The average in Eq. (2.2) requires integrating over both the Boltzmann factor $\exp{\left( -\sca{$\beta$H}\right)}$ and $\exp{\left( \sca{hM}\right)}$ factor, but as the field \sca{h} strengthens, the $\sca{hM}$ term will eventually overcome the contributions of the Hamiltonian $\sca{H}$.  Thus, in that limit, the molecular degrees of freedom become independent, and Eqs. (2.2) and (2.3) depends on just a single-molecule average
\begin{equation}
    \sca{$\lambda$}_{\text{strong-field}}(\sca{h}) = \ln{\langle \exp{\left( \sca{hx} \right)}\rangle} \tag{3.4}\label{eq:3.4}
\end{equation}
\begin{equation}
    \sca{m} = \frac{\langle x \exp{\left( \sca{h*x} \right)}\rangle}{\langle \exp{\left( \sca{h*x} \right) \rangle}} \tag{3.5}\label{eq:3.5}
\end{equation}
Intermolecular correlations are manifestly unimportant in this (perhaps physically unachievable) limit, but the average in Eq. (3.4) embodies the full non-linearity of each molecule’s contribution and might therefore be able to capture the large-fluctuation regime in quantitative detail. 
\parskip=1pt
\par It may be worth elaborating a bit on how this large-field dominance plays out in realistic liquid-crystal situations. Intermolecular potentials in liquids, at least in non-hydrogen-bonding cases, do have steep and sizeable repulsive core potentials, along with more slowly varying, and softer, attractive potentials. However, the potentially problematic large repulsive energies are effectively masked because the molecular-core/molecular-core overlaps they correspond to are configurations of measure zero at liquid temperatures. The key observation is that even with unbounded potentials, the field can always be chosen to be much larger than ($\beta$ times) the size of typical intermolecular potential energy fluctuations, as measured, say, by the square root of the excess heat capacity $\Delta C_v$ (which is proportional to the root-mean-square potential energy fluctuation). Susceptibilities such as the heat capacity do begin to diverge as the isotropic/nematic avoided critical point is approached. But we shall find that we can profitably apply the large-fluctuation methods of this paper well before the growth in the susceptibilities becomes problematic, something one can verify by looking at actual orientational susceptibility values for the model liquid crystal system we study in this paper (given in the Supplementary Material).

The basic strategy we will pursue in the remainder of the paper is to devise the minimal, symmetry-allowed, interpolation of the  equation of state that respects both the weak and the strong-field limits.  Somewhat surprisingly for a many-body problem, the key will be accurately building in the single-molecule nonlinearity.  The linear, low-field-limit, equation of state one gets in almost every problem, Eq. (3.2), is basically Hooke’s law -- the stress ($\sca{h}$) is proportional to the strain ($\sca{$\delta$m}$) -- but at large fields, problem-specific inelastic (nonlinear) behavior sets in.  For the orientational-order problems that we will pursue, for example, the single molecule quantities $\sca{x}$ in Eq. (3.3) inevitably saturate at some maximum value, which entropic considerations dictate can only be achieved if the corresponding field diverges.  In other words, our  equations of state will invariably have a pole at $\sca{m} = \sca{m}_{max}$.  Explicitly incorporating that pole (and the leading corrections to it) will turn out to be enough to successfully predict large fluctuation behavior.

\section{\label{sec:level1}NEMATIC LIQUID CRYSTALLINE ORDER IN THE ISOTROPIC PHASE}

\subsection{General considerations}

The most general way to measure nematic order (the extent to which a liquid of molecules is lined up along a particular, but unspecified, axis) is to construct a second-rank order-parameter tensor $\matr{Q}$.\cite{degennes} If we denote the orientation unit vector specifying the direction of the $\sca{j}$-th molecule by $\unitvector{\Omega}_j$, then in $\sca{d}$-dimensions the tensor takes the form of a symmetric, traceless  matrix
\begin{equation}
    \matr{Q} = \frac{1}{\sca{N}} \displaystyle \sum_{j=1}^{N} \matr{q}_j, \quad \matr{q}_j = \frac{\sca{d} \unitvector{\Omega}_j\unitvector{\Omega}_j - \matr{1}}{\sca{d} - 1} \tag{4.1}\label{eq:4.1}
\end{equation}
with $\matr{1}$ the unit matrix.  The largest eigenvalue of this matrix $\sca{s}$ is the desired order parameter.  Since the corresponding eigenvector is the director unit vector $\unitvector{n}$ (the preferred axis), that eigenvalue
\begin{equation}
    \sca{s} = \frac{1}{\sca{N}} \displaystyle \sum_{j=1}^{N} \frac{\sca{d} \left(\unitvector{\Omega}_j \cdot \unitvector{n}\right)^2 - 1}{\sca{d} - 1} = \frac{1}{\sca{N}} \displaystyle \sum_{j=1}^{N} \frac{\sca{d} \: (\cos^2{\sca{$\theta$}_j}) - 1}{\sca{d} - 1}, \quad \left(0 \le \sca{s} \le 1 \right) \tag{4.2}\label{eq:4.2}
\end{equation}
(with $\sca{$\theta$}_j$ the angle the $\sca{j}$-th molecule makes with the director) is a revealing measure of where the system falls between the extremes of being perfectly isotropic $\langle \cos^2{\sca{$\theta$}} \rangle = \sca{d}^{-1} \Rightarrow \sca{s} = 0$ and completely ordered ($\langle \cos^2{\sca{$\theta$}} \rangle = 1 \Rightarrow \sca{s} = 1$).

For a macroscopic system in the isotropic phase, there is, of course, no orientational order, so the probability distribution of $\sca{s}$ values,  $\sca{p(s)}$, is simply a delta function at $\sca{s} = 0$.  However, any finite simulation will have a range of order parameter fluctuations, and hence a nontrivial distribution function. \cite{frenkel, Eppenga1985}  Indeed, these fluctuations are just what we wish to explore.  So, what we need to know, in practice, is how this distribution scales with increasing $\sca{N}$ in the isotropic phase.  Let us look at what we can say, in turn, about the nematic order expected in finite two-dimensional and three-dimensional isotropic liquid crystalline systems.

\subsection{Two-dimensional liquid crystals}

Since the order parameter tensor $\matr{Q}$ is symmetric and traceless, there are only two independent quantities needed to specify it in two dimensions. \cite{Eppenga1985}
\begin{equation}
    \matr{Q} = \begin{pmatrix}
    \ Q_1 & Q_2 \\
    \ Q_2 & -Q_1 \\
    \end{pmatrix} \tag{4.3a}\label{eq:4.3a}
\end{equation}
In particular, denoting the laboratory-frame axes by $\sca{x}$ and $\sca{y}$ lets us write
\begin{equation}
\begin{aligned}
Q_1 &= \frac{1}{N} \sum_{j=1}^{N} q_{j}^{(xx)}, \quad  q_{j}^{(xx)} = 2 \cos^2{\theta}_j - 1 = \cos{2\theta_j}\\
Q_2 &= \frac{1}{N} \sum_{j=1}^{N} q_{j}^{(xy)}, \quad q_{j}^{(xy)} = 2\cos{\theta_j} \sin{\theta_j} = \sin{2\theta_j} 
\end{aligned} \tag{4.3b}\label{eq:4.3b}
\end{equation}
with $\theta_j$ now the angle of the $\sca{j}$-th molecule from the $\sca{x}$ axis (rather than from as-of-yet-unknown director). This structure means that the information needed to specify the $\matr{Q}$ tensor can be expressed as the components of a two-dimensional vector, \footnote{In anticipation of how the formalism will play out in three dimensions, we point out that the “vectors” here are not the usual 2-$\sca{d}$ vectors in a Cartesian-coordinate basis; because our components involve trigonometric functions of double angles, rotating the coordinate axes does not make the components of our vectors transform in the way one would expect from Cartesian coordinates.}
\begin{equation}
    \vect{Q} = \begin{pmatrix}
    \ Q_1 \\
    \ Q_2 \\
    \end{pmatrix} = \frac{1}{N} \sum_{j=1}^{N} \vect{q}_{j}, \quad \vect{q}_{j} = \begin{pmatrix}
    \ \cos{2\theta_j} \\
    \ \sin{2\theta_j} \\
    \end{pmatrix} \tag{4.4}\label{eq:4.4}
\end{equation}
the magnitude of which, $\sca{Q}$, is actually the desired order parameter. \footnote{Any traceless $2 \times 2$ matrix $\matr{Q}$ has eigenvalues $(s, -s)$, so for us, the sum of the squares of the eigenvalues is $\left(-s\right)^2+\left(s\right)^2=2s^2=Tr\left(\matr{Q}^2\right) =2\left(Q_{1}^2+Q_{2}^2\right)$.}
\begin{equation}
    s = Q \equiv |\vect{Q}| = \sqrt{Q_1^2 + Q_2^2} = \frac{1}{N} \sqrt{\displaystyle \sum_{j,k=1}^{N} \cos{2(\theta_j - \theta_k)}} \tag{4.5}\label{eq:4.5}
\end{equation}

This formulation makes applying the vector version of large-deviation formalism, Eqs. (2.6) - (2.8), straightforward. The extensive quantity we are targeting is $\sca{N} \vect{Q}$, so the desired probability density of the $\vect{Q}$ vector is
\begin{equation}
    P(\vect{Q}) \sim \exp{\left( -N I(\vect{Q})\right)}, \quad \int d\vect{Q} P(\vect{Q}) = \int dQ_1 \int dQ_2 P(\vect{Q}) = 1 \tag{4.6}\label{eq:4.6}
\end{equation}
with the equation of state connecting $\vect{Q}$ to the the conjugate vector field $\vect{h}$.
\begin{equation} 
\exp{\left( N \lambda(\vect{h}) \right)} = \langle \exp{(N \vect{Q} \cdot \vect{h})} \rangle, \quad \vect{Q} = \nabla_{\vect{h}} \lambda(\vect{h}) \tag{4.7}\label{eq:4.7}
\end{equation}
determining the rate function $I(\vect{Q})$
\begin{equation}
    I(\vect{Q}) = \int_{\vect{0}}^{\vect{Q}} \vect{h}(\vect{Q}') \cdot d\vect{Q}' \tag{4.8}\label{eq:4.8}
\end{equation}

Now consider the limiting behaviors of the equation of state when the liquid is isotropic. In the weak-field limit, we obtain central-limit-theorem behavior that is the 2-$d$ vector generalization of Eq. (3.1). As shown in Appendix A, isotropy implies the $2 \times 2$ matrix of order-parameter-vector fluctuations $\delta \vect{Q} = \vect{Q} - \langle \vect{Q} \rangle$, proportional to the two-dimensional orientational susceptibility $\chi$,
\begin{equation}
    \lambda_{\text{weak-field}}(\vect{h}) = \frac{1}{2} N \displaystyle \vect{h} \cdot \langle \delta \vect{Q} \delta \vect{Q} \rangle \cdot \vect{h} = \frac{\chi}{4} \vect{h} \cdot \vect{h} = \frac{\chi}{4} h^2 \tag{4.9}\label{eq:4.9}
\end{equation}
\begin{equation}
    \vect{Q} = Q \unitvector{h}, \quad Q = \frac{\chi}{2} h \tag{4.10}\label{eq:4.10}
\end{equation}
\begin{equation}
\chi = N\left \langle s^2 \right \rangle = \left \langle \frac{1}{N} \displaystyle \sum_{j,k=1}^{N}\cos{2(\theta_j - \theta_k)} \right \rangle \tag{4.11}\label{eq:4.11}
\end{equation}
with $\unitvector{h}$ a unit vector in the direction of the $\vect{h}$ field and $h$ the magnitude of the field. In the strong-field limit, we find (since Eq. (4.4) is the analog of Eq. (3.3)), that the independent-molecule expression, Eq. (3.4) becomes
\begin{equation}
    \lambda_{\text{strong-field}}(\vect{h}) = \ln{\left \langle \exp{(\vect{q} \cdot \vect{h})} \right \rangle} = \ln{\left(I_0(h) \right)} \tag{4.12}\label{eq:4.12}
\end{equation}
\begin{equation}
    \vect{Q} = Q \unitvector{h}, \quad Q = \frac{I_1(h)}{I_0(h)} \tag{4.13}\label{eq:4.13}
\end{equation}
with $I_0$ and $I_1$ zeroth- and first-order modified Bessel functions.

Equations (4.10) and (4.13) tell us the limiting behaviors of the $Q(h)$ equation of
state. Using the asymptotic behavior of the Bessel functions, we can write
\begin{equation*}
Q =
\begin{cases}
\frac{\chi}{2} h, \quad h \rightarrow 0\\
1 - \frac{1}{2h} - \frac{1}{8h^2} + ..., \quad h \rightarrow \infty
\end{cases} 
\end{equation*}
but in order to use Eq. (4.8) to evaluate the rate function, it is more convenient to use what we know to devise an interpolation formula for the inverse expression, $h(Q)$.
\begin{equation}
h(Q)=
\begin{cases}
\frac{2}{\chi} Q, \quad Q \rightarrow 0\\
\frac{1}{2(1-Q)} + \frac{1}{4} + \mathcal{O}\left( 1-Q \right), \quad Q \rightarrow 1
\end{cases} \tag{4.14}\label{eq:4.14}
\end{equation}

Perhaps the simplest way to represent Eq. (4.14) while doing justice to the pole at $Q = 1$ is to write a Padé approximant. 
\cite{Jedynak, Cohen, press1986}
But in doing so, it is helpful to keep in mind two considerations. The first is symmetry. The probability distribution of the $(s, -s)$
eigenvalues of the $\matr{Q}$ matrix should obviously be invariant to the eigenvalue sign, so Eq. (4.5) tells us that $I(Q)$, the rate function for the probability distribution of the $\vect{Q}$ vector,
must involve only even powers of $Q$. That, in turn, requires (via Eq. (4.8)) that $h(Q)$ must be odd in $Q$.

A second consideration is what the leading-order effects of intermolecular
correlations are on the $Q \rightarrow 1$ limit. We have treated that limit in Eq. (4.14) as one with no correlations, just as we discussed in Sec. III, giving us a constant as the leading correction to the pole at $Q = 1$. But to leading order, the actual effect of adding correlations will merely be to change this constant: the correlations simply add a net shift of the applied field h reflecting the potential energy each molecule feels in a hypothetically fully-orientationally-ordered liquid crystal. We can therefore rewrite Eq. (4.14) as
\begin{equation}
h(Q)=
\begin{cases}
\frac{2}{\chi} Q, \quad Q \rightarrow 0\\
\frac{1}{2(1-Q)} + \eta + \mathcal{O}\left( 1-Q \right), \quad Q \rightarrow 1
\end{cases} \tag{4.15}\label{eq:4.15}
\end{equation}
and view the quantity $\eta$ the way we view $\chi$, as a constant to be determined by the simulation.

The very simplest Padé that is odd in $Q$ and correctly reproduces the limits in Eq. (4.15) looks elaborate
\begin{equation}
h(Q) = \frac{2}{\chi} \frac{Q + A Q^3 + B Q^5}{1 - Q^2} \tag{4.16}\label{eq:4.16}
\end{equation}
\begin{equation}
A = -2 + \left(\frac{\chi}{2}\right) \left( \eta + \frac{9}{4} \right), \quad B = 1 - \left(\frac{\chi}{2}\right) \left( \eta  + \frac{5}{4} \right) \tag{4.17}\label{eq:4.17}
\end{equation}
but readers can convince themselves that Eq. (4.16) is, in fact, the minimal form that will meet all of our criteria. \footnote{The Padé approximant given by Eq. (4.16) is the simplest possible form that is odd in
$Q$, possesses the lowest possible order in the denominator, and allows for arbitrary values of the two parameters $\chi$ and $\eta$. Were we to remove the requirement of incorporating the correct value of $\eta$ (the leading order correction to the perfect order behavior), we could replace the quintic polynomial in the numerator of this Padé with a cubic polynomial, which would lead to a simpler looking rate function as well. But, as we show in the Supplementary Material, that would not produce an improvement over a straightforward application of the central-limit-theorem (CLT) theory to the orientational order probability density. In fact, it would often be worse than the CLT. Moreover, the rate function would only be superficially simpler: its Landau expansion would still have $Q^4$ and higher-order contributions coming from an expansion of the $\ln{(1-Q^2)}$ term.} The probability distribution, Eq. (2.6), can then be evaluated analytically from the reversible work integral, Eq. (4.8).
\begin{equation}
I(\mathbf{Q}) = \left(\frac{\delta}{2}\right) Q^2 - \left(\frac{B}{2\chi}\right) Q^4 - \frac{1}{2}\ln{(1-Q^2)} \tag{4.18a}\label{eq:4.18a}
\end{equation} 
\begin{equation}
P(\mathbf{Q}) \sim \left( 1-Q^2 \right)^{\frac{N}{2}} \exp{ \left \{ -N \left[ \left(\frac{\delta}{2}\right) Q^2 - \left(\frac{1 + \delta}{4}\right)B Q^4 \right] \right \} } \tag{4.18b}\label{eq:4.18b}
\end{equation}
For notational simplicity here, we have introduced the susceptibility index $\delta = \frac{2}{\chi} - 1, -1 \le \delta \le 1$ which describes behavior ranging from fully correlated $\chi=N \Rightarrow \delta = -1$ in the large $N$ limit) to uncorrelated ($\chi=1 \Rightarrow \delta = 1$). The final goal, $p(s)$, the probability
density of the order parameter $s = Q$, then falls directly out of Eq. (4.6)
\begin{equation}
p(s) = 2\pi s \left[ P(\mathbf{Q}) \right]_{Q=|\mathbf{Q}|=s} \tag{4.19}\label{eq:4.19}
\end{equation}

The structure of Eq. (4.18a) may seem unfamiliar, but when $Q$ is small, the rate function is just what one would expect from the Landau free energy in the isotropic phase. Under these circumstances, we can expand the log term to find
\begin{equation*}
    I(\mathbf{Q}) = \frac{a}{2} Q^2 + \frac{b}{4} Q^4 + ... \quad \left( a=\frac{2}{\chi}, \quad b=\frac{5}{4}-\delta + \eta \right)
\end{equation*}
with the quadratic term the central-limit-theorem result, and $a$ and $b$ both positive. What is new is what happens when $Q$ is not small. As the fluctuations grow, and, in particular, when as $Q$ approaches its maximum value of 1, we find that the effective free energy now starts to feel a logarithmic singularity.

\subsection{Three dimensional liquid crystals}

In three dimensions, the order parameter tensor, Eq. (4.1), is made up of five independent elements, each of which is a sum of single-molecule contributions
\begin{equation}
    \matr{Q} = \begin{pmatrix}
    \ Q_1 & Q_4 & Q_5 \\
    \ Q_4 & Q_2 & Q_6\\
    \ Q_5 & Q_6 & Q_3\\
    \end{pmatrix}, \quad Tr(\matr{Q}) = Q_1 + Q_2 + Q_3 = 0 \tag{4.20}\label{eq:4.20}
\end{equation}
\begin{equation}
    Q_{\alpha} = \frac{1}{\sca{N}} \displaystyle \sum_{j=1}^{N} q_{j\alpha}, \quad \left(\alpha = 1,2,4,5,6\right) \quad \text{for example} \tag{4.21}\label{eq:4.21}
\end{equation}
These elements could be construed to be elements of a $\vect{Q}$ vector in Eqs. (4.6)-(4.8), much as we did for the two elements of the two-dimensional case. However, it is easier to see the essential symmetries of this problem if we realize that the five elements of each $\matr{q}_{\sca{j}}$ entering the full $\matr{Q}$ tensor are just linear combinations of the five $\sca{l} = 2$ spherical harmonics $\sca{Y}_{\sca{lm}}(\unitvector{\Omega}_j), \: m = -2, ..., 2$. Better yet, they can be viewed as linear combinations of the equivalent five $\sca{l} = 2$ tesseral (real) spherical harmonics \cite{blanco} $\sca{R}_{\sca{lm}}(\unitvector{\Omega}_j)$ . The translation between the individual $Q_{\alpha}$ and the ($\sca{l} = 2$) $\sca{R}_m$ is given in Appendix B.

This equivalence suggests we choose our extensive order-parameter vectors to be (first rank) $l = 2$ spherical tensors $\vect{R} = \{R_m; m = - 2,...,2\}$
\begin{equation}
    R_m = \frac{1}{N} \displaystyle \sum_{j=1}^{N} r_m(\hat{\Omega}_j), \quad \left( m=-2,...2 \right) \tag{4.22}\label{eq:4.22}
\end{equation}
and likewise for the conjugate field $\vect{h} = \{h_m; m = - 2,...,2\}$ , with
\begin{equation}
    \vect{h} \cdot \vect{R} = \displaystyle \sum_{m=-2}^{2} (h_m) (R_m), \quad h^2 = \vect{h} \cdot \vect{h} = \displaystyle \sum_{m=-2}^{2} (h_m)^2 \tag{4.23}\label{eq:4.23}
\end{equation}
As we show in Appendix B, the sum of the squares of eigenvalues of the order-parameter tensor in this notation is given by
\begin{equation}
    Tr\left(\vect{Q}^2\right) = \frac{6\pi}{5} R^2 = \frac{6\pi}{5} \vect{R} \cdot \vect{R} = \frac{6\pi}{5} \displaystyle \sum_{m=-2}^{2} (R_m)^2 \tag{4.24}\label{eq:4.24}
\end{equation}

When written in this language, our cumulant generating function, Eq. (4.7), becomes
\begin{equation}
    \exp{(N \lambda(\vect{h}))} = \langle \exp{\left(N \vect{h} \cdot \vect{R} \right)} \rangle = \left \langle \exp{\left( \displaystyle \sum_{j=1}^{N} \displaystyle \sum_{m=-2}^{2} h_m r_m(\hat{\Omega}_j) \right)} \right \rangle \tag{4.25}\label{eq:4.25}
\end{equation}
Let us again consider the limiting behaviors in the situation when the macroscopic system is isotropic. In the weak-field limit, cumulant expansion through second order again generates the central-limit theorem predictions. Under isotropic conditions, the 5 components of the field $\vect{h}$ can be taken to be identical:
\begin{equation*}
    h_m^2 = \frac{1}{5} h^2, \quad (m=-2,...,2) 
\end{equation*}
As shown in Appendix B, this isotropy implies
\begin{equation}
    \lambda(\vect{h}) = \frac{1}{2N} \displaystyle \sum_{j,k=1}^{N} \displaystyle \sum_{m,m'=-2}^{2} h_m \left \langle r_m(\hat{\Omega}_j) r_{m'}(\hat{\Omega}_k)  \right \rangle  h_{m'} = \frac{\chi}{8\pi} h^2 \tag{4.26}\label{eq:4.26}
\end{equation}
with $\chi$ now the relevant three-dimensional orientational susceptibility
\begin{equation}
    \chi = \left \langle \displaystyle \frac{1}{N} \sum_{j,k=1}^{N} P_2(\hat{\Omega}_j \cdot \hat{\Omega}_k) \right \rangle \tag{4.27}\label{eq:4.27}
\end{equation}
making the corresponding equation of state (via Eq. (4.7)),
\begin{equation}
    \vect{R} = \frac{\chi}{4\pi} \vect{h} \Rightarrow R = \frac{\chi}{4\pi} h \tag{4.28}\label{eq:4.28}
\end{equation}

In the opposite, strong-field, limit, the field h can be regarded as being solely along the $z$-axis, meaning that only the $m = 0$ component contributes
\begin{equation*}
    h_m = h \delta_{m,0}
\end{equation*}
and the cumulant generating function has only single-molecule contributions. Thus
\begin{equation}
    \lambda(\mathbf{h}) = \ln{\left \langle \exp{(h r_0)} \right \rangle} = \ln{F(z)} -\ln{z} + \frac{2}{3} z^2 \tag{4.29}\label{eq:4.29}
\end{equation}
\begin{equation*}
    z^2 = h \sqrt{\frac{45}{16\pi}}, \quad F(z) = \exp{(-z^2)} \int_{0}^{z} \exp{(x^2)} dx
\end{equation*}
(that is, $F(z)$ is Dawson’s function ),\cite{wikipedia_dawson_function} making the equation of state in this limit
\begin{equation}
    R = \sqrt{\frac{45}{64\pi}} \left( \frac{1}{zF(z)} - \frac{1}{z^2} - \frac{2}{3} \right) \tag{4.30}\label{eq:4.30}
\end{equation}

The appearance of the results can be simplified somewhat if we introduce a rescaled order parameter $r$ and field $H$ which preserves the product of the original work variables: $r H = R h$.
\begin{equation}
    r = \sqrt{\frac{4\pi}{5}} R, \quad H = \sqrt{\frac{5}{4\pi}} h \tag{4.31}\label{eq:4.31}
\end{equation}
Using the asymptotic behavior of Dawson’s function, \cite{wikipedia_dawson_function} then tells us that the extreme limits of our $r(H)$ equation of state are
\begin{equation*}
    r = \begin{cases}
    \frac{\chi}{5} H, \quad H \rightarrow 0 \\ 
    1 - \frac{1}{H} - \frac{1}{3H^2} + ..., \quad H \rightarrow \infty
\end{cases}
\end{equation*}
or, inverting to get the $H(r)$ equation of state
\begin{equation}H(r) =
\begin{cases}
    \frac{5}{\chi} r, \quad r \rightarrow 0 \\ 
    \frac{1}{1-r} + \eta + \mathcal{O}(1-r), \quad r \rightarrow 1
\end{cases} \tag{4.32}\label{eq:4.32}
\end{equation}
where, as we did in the two-dimensional case, we introduce a yet-to-be-determined constant $\eta$ that includes the contribution from the uncorrelated result (here $\frac{1}{3}$) plus the leading-order effects of intermolecular correlations.

Both of these limits can be effectively combined, as we did before, by introducing
a Padé approximant. Since we no longer have any $+/-$ symmetry, the minimal approximant is simply
\begin{equation}
    H(r) = \frac{5}{\chi}  \frac{r + Ar^2 + Br^3}{1 - r} \tag{4.33}\label{eq:4.33}
\end{equation}
\begin{equation} 
    A = -2 + \left( \frac{\chi}{5} \right) \left(\eta + 3 \right), \quad 
    B = 1 - \left( \frac{\chi}{5} \right) \left(\eta + 2 \right) \tag{4.34}\label{eq:4.34}
\end{equation}
which leads to the reversible work rate function
\begin{equation}
\begin{aligned} 
    I(r) &= \int_{0}^{R} h(R') dR' = \int_{0}^{r} H(r') dr' \\
    &= -\ln{(1-r)} - r + \frac{1}{2} \left(\frac{5}{\chi} - 1\right) r^2 +\frac{1}{3} \left(\eta + 2 - \frac{5}{\chi}\right) r^3 
\end{aligned} \tag{4.35}\label{eq:4.35}
\end{equation}

Equation (4.35) tells us the desired probability density of the $\vect{r} = \sqrt{\frac{4\pi}{5}} \vect{R}$ order parameter vector, $P(\vect{r}) \sim \exp{\left(-NI(\vect{r}) \right)}$ , which we have defined so that,
\begin{equation*}
    \int d\vect{r} P(\vect{r}) = 1
\end{equation*}
The answer we really want, though, is the probability distribution of the largest eigenvalue of the order parameter matrix – and we are not as fortunate as in the 2-$d$ case, where that eigenvalue turned out to be just the magnitude of the order parameter vector. Nonetheless, it turns out that we can finish off the calculation just by using some results from random-matrix theory.

We observe, first, that $P(r)$ depends only on the magnitude $r$ and remember from Eq. (4.24) and (4.31) that the sum of the squares of the three eigenvalues $(s_1, s_2, s_3)$ is
\begin{equation}
    s_1^2 + s_2^2 + s_3^2 = Tr\left(\matr{Q}^2\right) = \frac{6\pi}{5} R^2 = \frac{3}{2} r^2 \tag{4.36}\label{eq:4.36}
\end{equation}
So, our rate function depends only on the three eigenvalues. To obtain the probability distribution of those eigenvalues, then, we simply need to find the Jacobian, $J_1$, of the transformations from $\vect{r}$, the spherical tensor representation of the order-parameter matrix $\matr{Q}$, to $\vect{Q}$, the independent Cartesian matrix elements of $\matr{Q}$, and the Jacobian $J_2$ which takes $\vect{Q}$ into the eigenvalues of $\matr{Q}$. Inasmuch as the sum of the eigenvalues has to vanish, \footnote{Since $\matr{Q}$ is traceless, its eigenvalue sum is always zero.}
\begin{equation*}
    \int d\vect{r} = \int d\vect{Q} \:\:(J_1) = \int \int \int ds_1 \: ds_2 \: ds_3 \:\:(J_1 J_2) \:\: \delta(s_1 + s_2 + s_3)
\end{equation*}
The first Jacobian is easily shown to be a constant, meaning that it can be subsumed into the overall normalization; the second is known from random matrix theory to be the Vandermonde determinant, $J_2 = (s_1 - s_2)(s_2 - s_3)(s_1 - s_3)$, provided the eigenvalues are ordered $s_1 \ge s_2 \ge s_3$. Thus the probability density of the largest eigenvalue $s$ is
\begin{equation}
\begin{aligned}
    p(s) &\sim \int_{-\infty}^{s} ds_2 \int_{-\infty}^{s_2} ds_3 \:\: \delta(s + s_2 + s_3) \:\: (s-s_2)(s-s_3)(s_2-s_3) \\
    &\times \exp{\left(-NI \left(r=\sqrt{\frac{2}{3} \left(s^2 + s_2^2 + s_3^2 \right)}\right) \right)} 
\end{aligned} \tag{4.37}\label{eq:4.37}
\end{equation}

The basic form of our rate function regarded as a function of the order parameter matrix, $I(\matr{Q})$, is again, an unfamiliar one, but if we expand the rate function, Eq. (4.35), in powers of $r$ and make use of Eq. (4.36), we see that, as in the 2$d$ case, the small order-parameter limit
\begin{equation}
    I(\matr{Q}) = \frac{1}{2} a \: Tr\left(\matr{Q}^2\right) + \frac{1}{3} b \: \left(Tr\left(\matr{Q}^2\right)\right)^{3/2} + \frac{1}{4} c \: \left(Tr\left(\matr{Q}^2\right)\right)^2 + ... \tag{4.38}\label{eq:4.38}
\end{equation}
\begin{equation*}
    a = \frac{2}{3} \frac{5}{\chi}, \quad b = \left(\frac{2}{3}\right)^{3/2} \left[3 + \eta - \frac{5}{\chi} \right], \quad c = \left(\frac{2}{3}\right)^2
\end{equation*}
is similar to the conventional Landau form of the free energy \cite{Gennes, chakrabarti2009, gramsbergen}
\begin{equation}
    I(\matr{Q}) = \frac{1}{2} A \: Tr\left(\matr{Q}^2\right) + \frac{1}{3} B \: Tr\left(\matr{Q}^3\right) + \frac{1}{4} C \: \left(Tr\left(\matr{Q}^2\right)\right)^2 + ... \tag{4.39}\label{eq:4.39}
\end{equation}
frequently used to explain both why liquid crystals have a first-order isotropic/nematic phase transition, and how that transition preempts a potential critical endpoint for the isotropic phase. (We will revisit the differences between these two versions of the free energy in the discussion.) The more interesting feature, though, is how the growth of the larger order parameter fluctuations is controlled not just by the Landau free energy, but by a logarithmic singularity, much as it is in the 2-$d$ situation.

We note, parenthetically, that the presence of these logarithmic singularities was also appreciated in the liquid-crystal free-energy calculations of Ilg and coworkers. \cite{Ilg2011, Ilg2012, Gupta2013, Luo2014}
In fact, they used the same basic elements of interpolating between weakly and strongly
ordered limits and of applying thermodynamic integration that we did. However, there
are some noticeable differences between our respective calculations. We discuss these in
Appendix C.

\section{\label{sec:level1}NUMERICAL RESULTS}

In this section we will compare our large-deviation-theory predictions for the
finite-size dependence of the nematic-order probability distribution with what we get
directly from simulating a model liquid crystal. Our theory carries out an interpolation
between situations with weak fluctuations (but potentially strong correlations) and those
with weak correlations (but potentially sizeable fluctuation), so we will also be
comparing our predictions with their central-limit-theorem counterparts (which are at
their best in the former case). In addition, to provide a baseline for our subsequent
calculations, we will examine how our theory would do if there were no intermolecular
interactions at all (the extreme limit of the latter case).

Consistent with the general discussion in Sec. 3, the \textit{central-limit theorem} (CLT)
predictions for our free energies as a function of order parameter are simply the reversible
work functions obtained by integrating the weak-field-limit equations of state. From Eqs.
(4.10) and (4.28), respectively, we obtain
\begin{equation}
    I_{CLT}^{2d}(Q) = \frac{Q^2}{\chi}, \quad I_{CLT}^{3d}(Q) = \frac{5r^2}{2\chi} \tag{5.1}\label{eq:5.1}
\end{equation}
Explicit expressions for the probability densities of the s order parameter (the largest
eigenvalue of the order parameter tensor $\matr{Q}$), can then be derived by the same techniques
described in Sec. 4 for the full large-deviation-theory (LDT) free energies. As we discuss
later, these results can also be written in the suggestive forms
\begin{equation}
    I_{CLT}^{2d}(\matr{Q}) = \left(\frac{1}{2\chi}\right) Tr\matr{Q}^2, \quad I_{CLT}^{3d}(\matr{Q}) = \left(\frac{5}{3\chi}\right) Tr\matr{Q}^2 \tag{5.2}\label{eq:5.2}
\end{equation}
emphasizing the essential identity of random matrix theory with central limit theorem
predictions for liquid crystals. \cite{zhao}

The two-dimensional numerical results we present in what follows are the predictions of Eq. (4.19) and either Eq. (4.18) (LDT) or Eq. (5.1) (CLT). The analogous three-dimensional results are from Eq. (4.37) and either Eq. (4.35) (LDT) or Eq. (5.1)
(CLT).

\subsection{The liquid crystal ideal gas}

The \textit{non-interacting limit} of a $d$-dimensional liquid crystal is determined by the
orientational statistics of a collection of $N$ independent $d$-dimensional unit vectors, a classic problem in probability theory. \cite{Dutka, lord} Moreover, in three dimensions, the orientational probability distribution for centrosymmetric non-interacting molecules has been shown to be given
accurately by simple random-matrix theory. \cite{doerr, zhao} In the context of this paper, though, this ideal limit simply corresponds to setting the susceptibility $\chi=1$, and the parameter $\eta=\frac{1}{4}$ in two dimensions, and $\eta=\frac{1}{3}$ in three dimensions.

\begin{figure}
\centering
\includegraphics[width=0.60\textwidth, height=0.55\textheight]{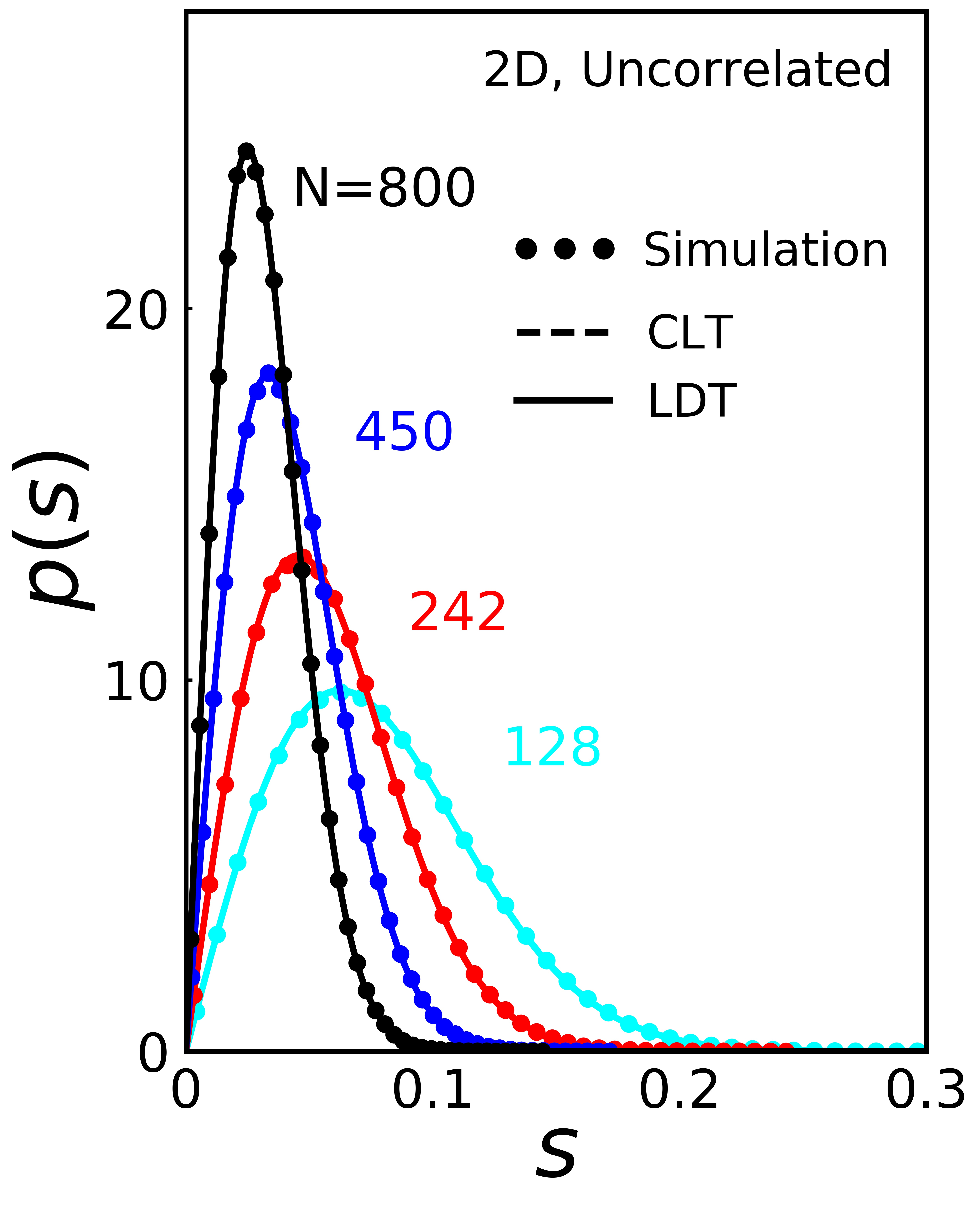}
\caption{Finite-size scaling results for the order-parameter probability distribution $p(s)$ of a \textit{two-dimensional non-interacting liquid crystal}. For each of the indicated values of the number of molecules $N$, molecular dynamics simulation results averaged over $10^6$ configurations (“simulation”) are compared with large deviation theory (LDT) and central-limit-theorem (CLT) predictions. The two theoretical predictions are indistinguishable on the scale of this figure.}
\label{fig:1}
\end{figure}

\begin{figure}
\centering
\includegraphics[width=0.60\textwidth, height=0.55\textheight]{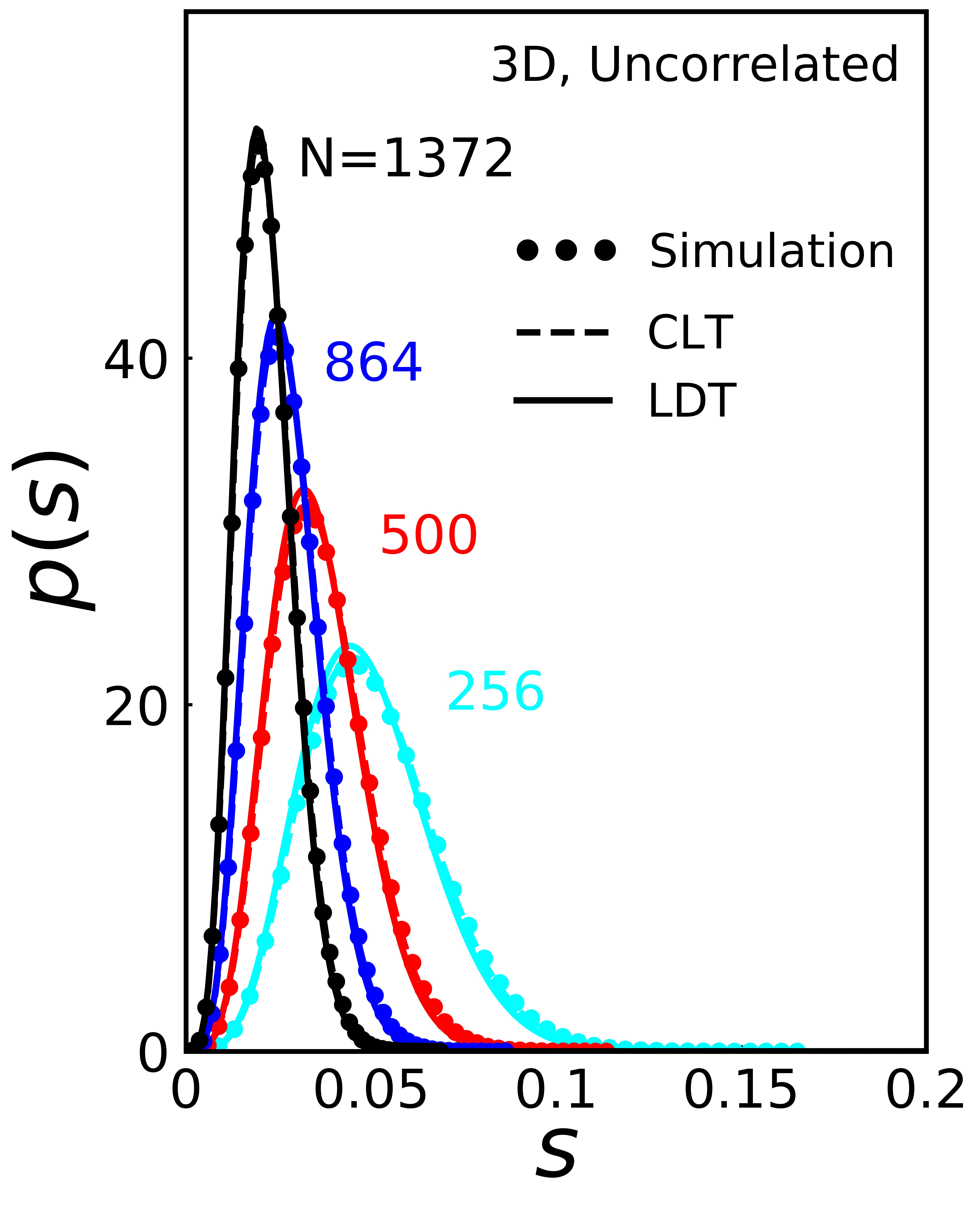}
\caption{Finite-size scaling results for the order-parameter probability distribution of a \textit{three-dimensional non-interacting liquid crystal} \cite{doerr}. The labeling and averaging in the figure is identical to that of FIG. ~\ref{fig:1}, and as with FIG. ~\ref{fig:1}, the two theoretical predictions are barely distinguishable on the scale of the figure.}
\label{fig:2}
\end{figure}

The large-deviation viewpoint here illustrates particularly clearly that the theory is just a large-$N$ motivated steepest descent approximation to the exact probability density. \footnote{In the language of Section 2, Eqs. (2.6)-(2.8), if the probability density we wanted was that of some $\vect{M}$ equal to a sum of $N$ unit vectors $\unitvector{\Omega}$, the textbook approach would write the exact answer as a Fourier transform, with respect to a vector $\vect{k} = i \: \vect{h}$ , of the generating function $\exp{(N\lambda(\vect{h}))}$ defined in Eq. (2.7). But, for independent unit vectors, $\exp{(N\lambda(\vect{h}))} = \left \langle \exp{\left(\vect{h} \cdot \unitvector{\Omega} \right)}\right \rangle^N$, meaning that as $N$ increases, the $\vect{k}$ integral is increasingly dominated by the steepest-descent $\vect{k}$ value, the one that makes the argument of the $\exp{(NI(\vect{m};\vect{h} = -i \vect{k}))}$ exponent stationary --- precisely the value that obeys the $(\vect{m}, \vect{h})$ relation Eq. (2.8)) - thereby demonstrating how large deviation theory becomes exact in the limit that $N$ is large. Similar arguments can be made for the probability distribution of an order parameter matrix $\matr{Q}$ equal to the sum of $N$ single-molecule matrices $\matr{q}$.} But the fundamental reliance on large $N$ does not mean that we need $N$ to be all that sizeable in practice. Indeed, as is evident in both two dimensions (Fig. \ref{fig:1}) and three dimensions (Fig. \ref{fig:2}), it is difficult to discern any deviations from the numerically exact simulation for any of the finite $N$ probability distributions on the scale of these figures. In particular, the finite-size scaling as $N$ increases is predicted quantitatively. However, we should emphasize that the potentially problematic large fluctuations that could derail this agreement are exceedingly rare in the absence of correlations. The near identity of both the simulated and large deviation results with the central-limit-theorem predictions bear out the fact that our large deviation theory, while accurate in this situation, is not especially necessary here.

\subsection{Fully interacting liquid crystals}

A more interesting test arises when we apply the same formalism to a realistic liquid crystal model. We choose to examine the Gay-Berne model \cite{gayberne}, a familiar generalization of the Lennard-Jones potential to anisotropic molecules, and one equally well adapted to either two \cite{Kundu, KUNDU2021113224, calamitic} or three \cite{demiguel, chakrabarti2007, jose2004a, jose2004b, jana2007a, jana2007b} dimensions. We choose our parameters to be the well-studied set \cite{demiguel, demiguel, chakrabarti2007, jose2004a, jose2004b, jana2007a, jana2007b}

\begin{equation*}
    \epsilon_{ss}/\epsilon_{ee} = 5, \quad \sigma_{\parallel}/\sigma_{\bot} = 3, \quad \mu=2, \quad \nu=1
\end{equation*}
where the first value is the ratio of the potential well depths for side-to-side and end-to-end orientations of a molecular pair, and the second is the ratio of the lengths of the major and minor axes of each molecule. The remaining two numbers are parameters controlling
the effect of those anisotropies on the potential.

One advantage of this parameter choice is that the associated phase diagram is sufficiently worked out that we can easily tune our thermodynamic conditions while remaining within the isotropic liquid range. In terms of reduced temperatures and densities defined by $T^* = kT/\epsilon_0, \rho^* = \rho \sigma_0^d$, with reference energy and length scales
\begin{equation*}
    \epsilon_0 = 3 \epsilon_{ee}, \quad \sigma_0 = \sigma_{\bot}
\end{equation*}
it turns out that we can choose the same $T^*=1.00$ in both $d = 2$ and $d = 3$, and watch the transition from weak to strong fluctuations just by increasing the value of $\rho^*$.

All of our simulations are carried out under periodic boundary conditions using microcanonical molecular dynamics with velocity-Verlet center-of-mass propagation and Fincham’s algorithm for molecular reorientation. \cite{allen1987computer} Our time step was $10^{-3} \tau_0$ (with $\tau = \left( m \sigma_0^2 / \epsilon_0 \right)^{1/2} $ the characteristic Lennard-Jones time), and data were collected only after a $200 \tau_0$ preparation regimen designed to gradually transform the system into the desired thermodynamic state, followed by a further free relaxation period of $200 \tau_0$ in which the system’s time evolution is allowed to proceed as directed by its Hamiltonian, without any outside interference. \footnote{The preparation regimen was based on the one used by Zhao and Stratt (Ref. 20). Relaxation processes slow noticeably as the isotropic/nematic transition is approached, but we know the scale of those relaxation times from collective reorientational time correlation functions computed in that paper, and potential-energy-component time correlation functions computed by one of the present authors. Comparison with these times imply that our relaxation protocol should generate a satisfactorily equilibrated set of liquid configurations at even the highest densities we studied in the present paper. Details are provided in, E. Mainas, Ph. D. dissertation (Brown University, 2023).} Liquid configurations were sampled every $10$ time steps.

Our large-deviation predictions depend on the simulations supplying the $\chi$ (susceptibility) and $\eta$ (leading correction to the saturated-order limit) parameters. We opted to recompute these parameters for every choice of density and number of
molecules we simulated. That allowed us to incorporate any subtle $N$-dependence that might have been present beyond the obvious $N$ dependence visible in equations such as Eq. (2.6). The calculation of the susceptibility, in particular, was a straightforward, outcome of the simulations via Eqs. (4.11) and (4.27) in two and three dimensions, respectively. The $\eta$ parameter, though, could have been obtained by a variety of means. To highlight the point that out theory can use sampling in high probability areas to inform its predictions for lower probability regions, we determined $\eta$ by requiring that the predicted average value of the order parameter $\langle s \rangle$, match that found by the simulations. The actual parameter values we use in the Gay-Berne figures that follow are listed in the Supplementary Material. As one can see from the tables shown there, the susceptibility $\chi$ systematically increases as the density increases and the systems get closer to their
avoided critical point, but neither parameter seems to have any significant $N$ dependence at given density. The predicted finite-size scaling of our large-deviation theory for liquid crystal order therefore apparently resides almost entirely in the $N$, and not the rate function $I$, in Eq. (2.6).

Carrying out the calculations for Gay-Berne liquid crystals shows that some of the same behavior we saw in the uncorrelated case recurs when we look deep in the isotropic phase. In two dimensions (Fig. 3), our large-deviation theory (LDT) predictions for the order-parameter probability distributions $p(s)$ are again indistinguishable from the
central-limit-theorem (CLT) predictions, and both are, again, virtually identical to the simulated distributions. In three dimensions, though, (Fig. 4) we see the first deviations between the CLT predictions and our LDT generalization. The differences are slight, but close examination shows that the latter converges noticeably more quickly to the simulated distributions as the number of molecules, $N$, increases.

\begin{figure}
\centering
\includegraphics[width=0.60\textwidth, height=0.55\textheight]{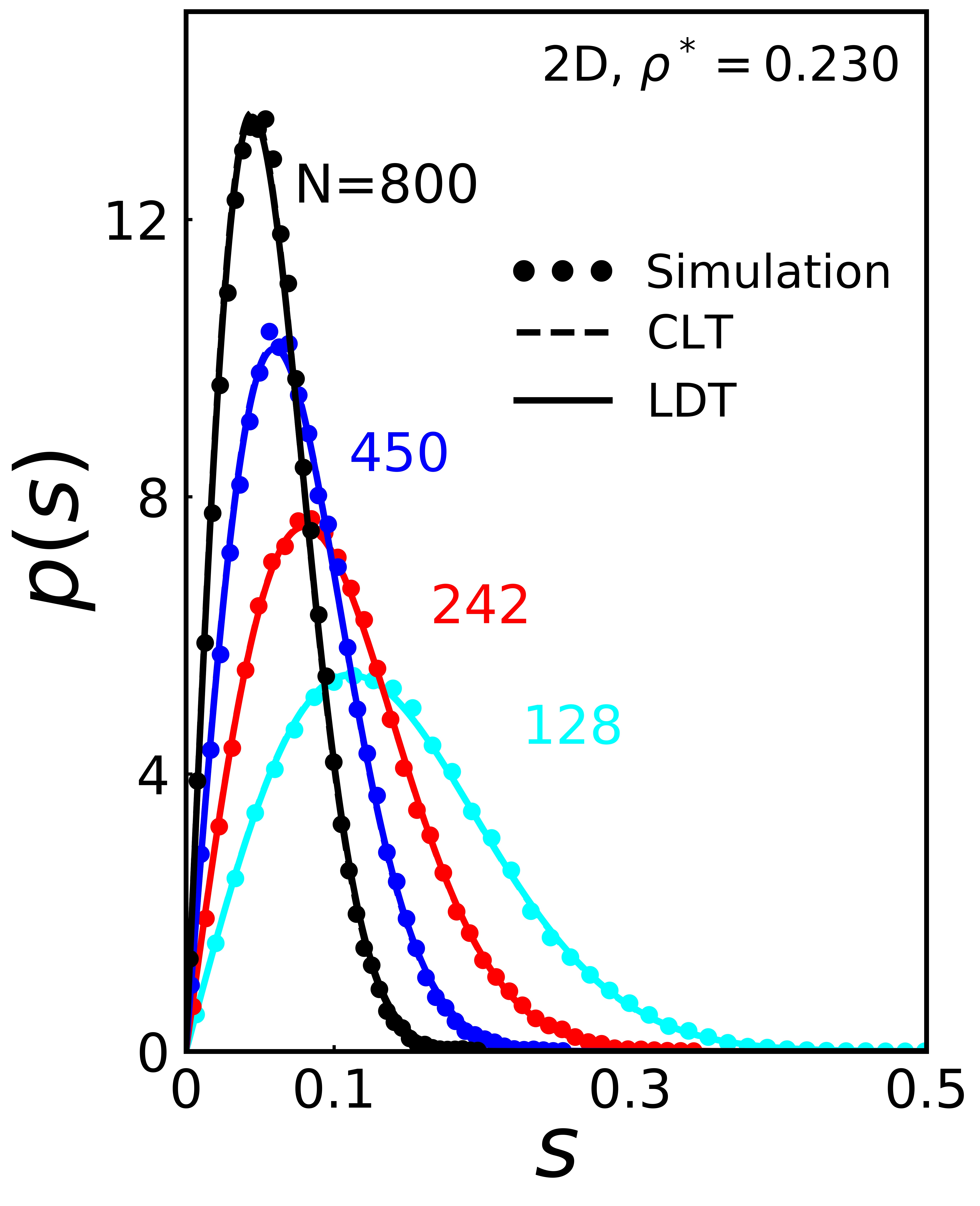}
\caption{Finite-size scaling results for the order-parameter probability distribution of a
\textit{two-dimensional Gay-Berne} liquid crystal deep in its isotropic phase ($T^*= 1.00$, $\rho^*= 0.230$). All molecular dynamics results are averaged over $3 \times 10^6$ configurations The labeling in this plot is identical to that in the previous figures, and as with those figures, the two theoretical predictions are indistinguishable on the scale of the figure.}
\label{fig:3}
\end{figure}

\begin{figure}
\centering
\includegraphics[width=0.60\textwidth, height=0.55\textheight]{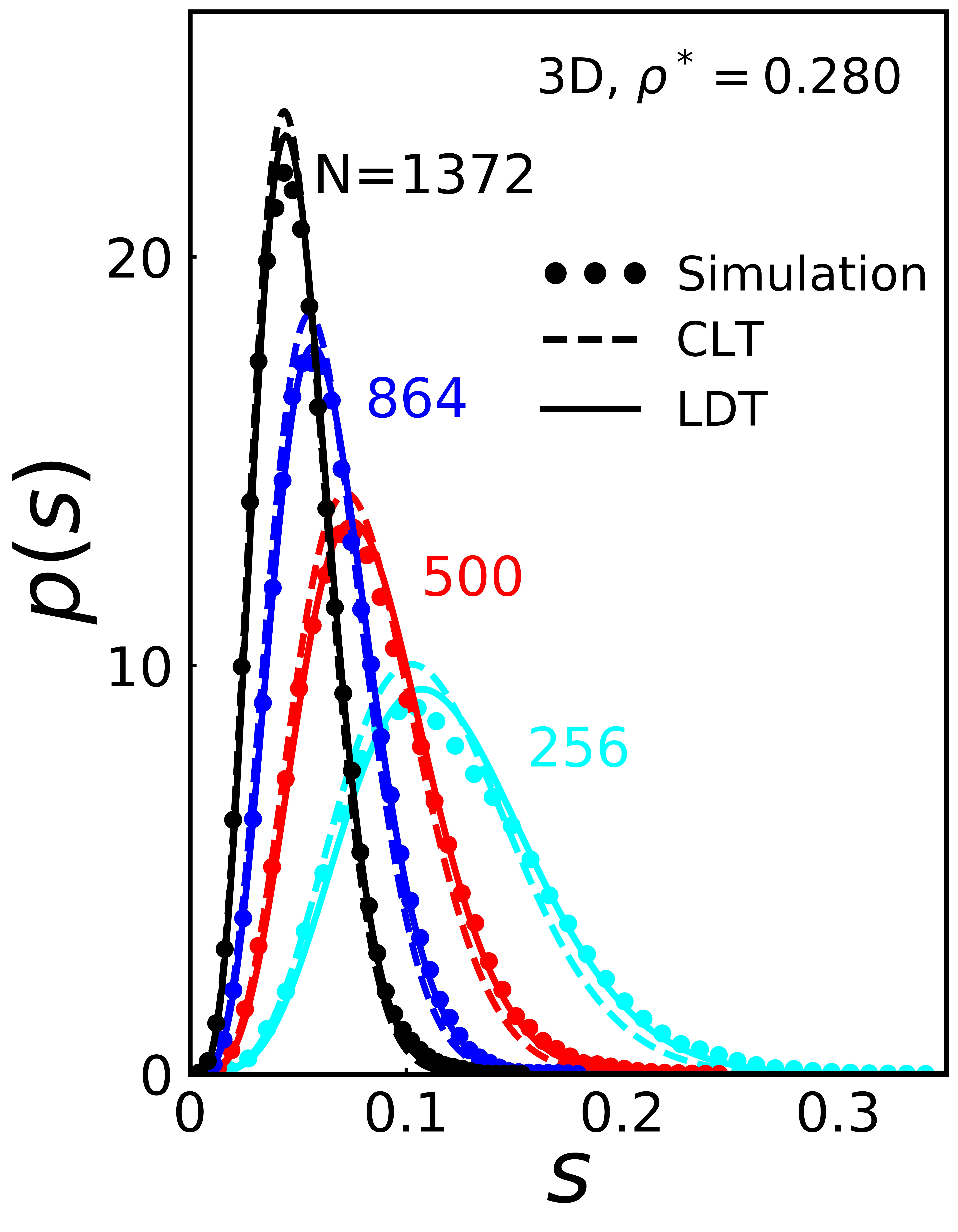}
\caption{Finite-size scaling results for the order-parameter probability distribution of a \textit{three-dimensional Gay-Berne} liquid crystal deep in its isotropic phase ($T^*= 1.00$, $\rho^*= 0.280$). The labeling and averaging in this plot is identical to that in FIG. ~\ref{fig:3}, but here the dashed-line (central-limit-theorem) predictions differ noticeably from the solid-line (large-deviation-theory) predictions at small $N$ values.}
\label{fig:4}
\end{figure}

These differences become much more pronounced when we move closer to the isotropic/nematic phase boundary. Both two-dimensional (Fig. 5) and three-dimensional
(Fig. 6) pre-transitional simulations now lead to order-parameter distributions that are fundamentally non-Gaussian (non-CLT), but the LDT predictions do a much better job of representing that non-Gaussian character, especially in the right-hand (large-fluctuation)
tail of the distribution. A look at the respective free energies ($\ln{p(s)}$ curves) in these regions, (Fig. 7), highlights just how much more closely our large-deviation theory tracks the actual shape of the true $p(s)$ tail.

\begin{figure}
\centering
\includegraphics[width=0.60\textwidth, height=0.55\textheight]{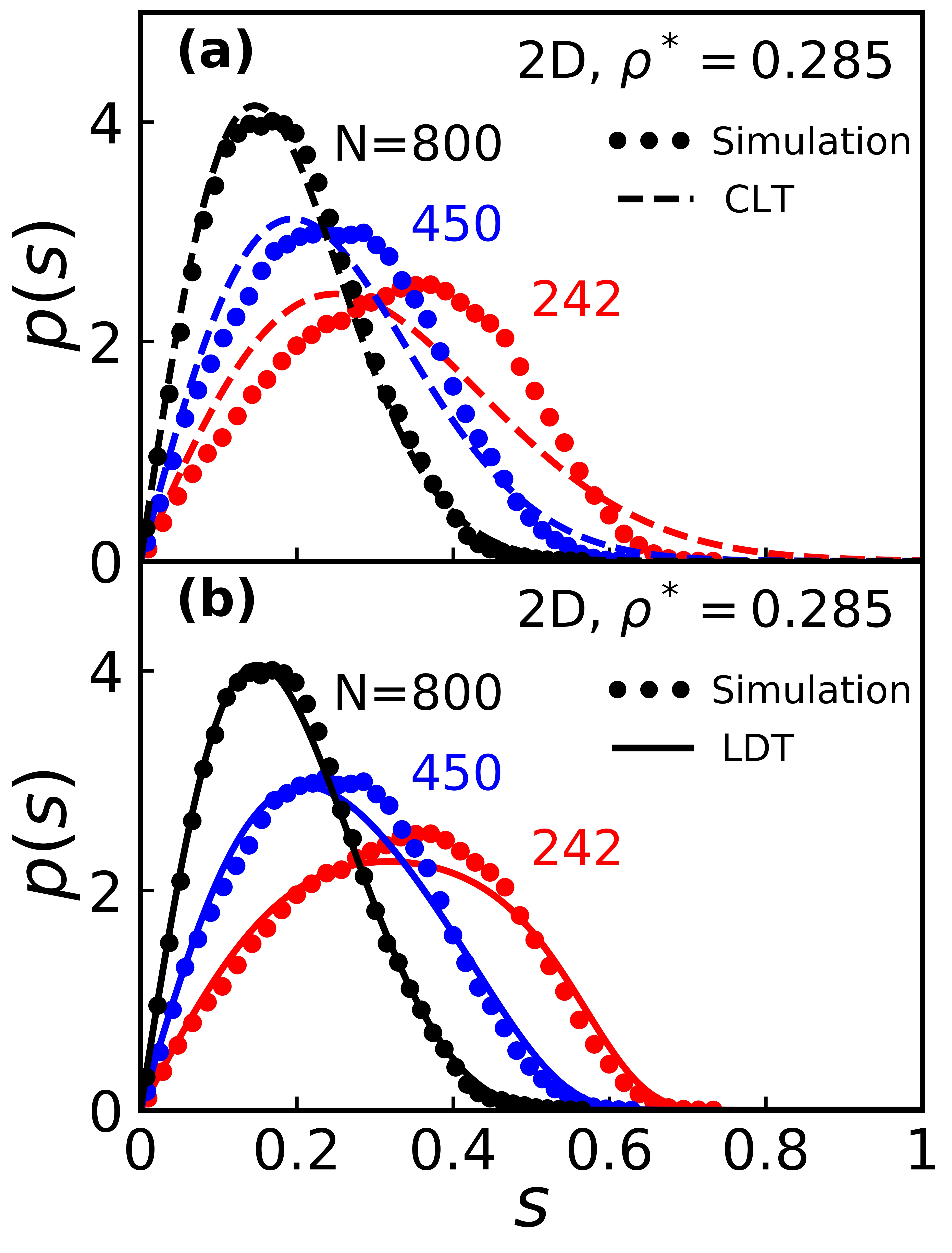}
\caption{Finite-size scaling results for the order-parameter probability distribution of an
isotropic \textit{two-dimensional Gay-Berne} liquid crystal in the pre-transitional region of the
phase diagram ($T^*= 1.00$, $\rho^*= 0.285$). All molecular dynamics results are averaged over $10^7$ configurations. The upper and lower panels compare these simulation results with the central-limit theorem, and large-deviation-theory predictions, respectively.}
\label{fig:5}
\end{figure}

\begin{figure}
\centering
\includegraphics[width=0.60\textwidth, height=0.55\textheight]{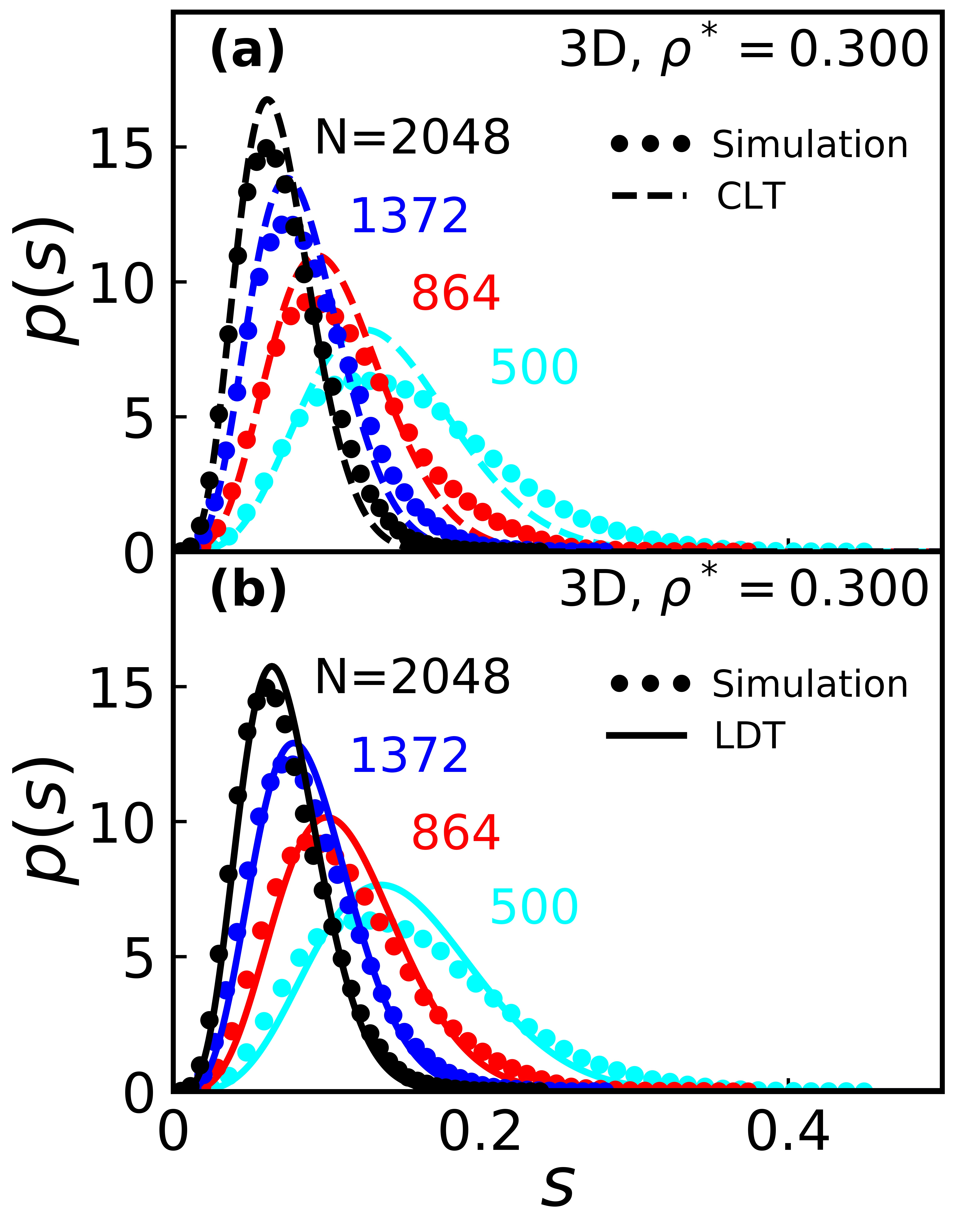}
\caption{Finite-size scaling results for the order-parameter probability distribution of an isotropic \textit{three-dimensional Gay-Berne} liquid crystal in the pre-transitional region of the phase diagram ($T^*= 1.00$, $\rho^*= 0.300$). All molecular dynamics results are averaged over $1.5 \times 10^7$ configurations. The upper and lower panels compare these simulation results with the central-limit theorem, and large-deviation-theory predictions, respectively.}
\label{fig:6}
\end{figure}

It might be well at this point to remind the reader that we are, in some sense comparing simulation results with simulation results. Our LDT curves are not, and are not meant to be, \textit{ab initio} statistical mechanical predictions; the theory curves unabashedly use some information extracted from simulation. However, the crucial feature is the way that our LDT formalism leverages that simulation information. Evaluating the $\chi$ and $\eta$ parameters relies mostly on information near the probability distribution maximum, but the theory nonetheless seems to be a reliable predictor of the low-probability large-fluctuation behavior seen in the distribution tails while
simultaneously capturing the global shape of the overall distribution.

That being said, we should also point out that there are limits to how large fluctuations can be and still be accommodated by the minimalist interpolation schemes proposed in this paper. If we venture closer still to the isotropic/nematic transition in three dimensions (Fig. 8), we find that for the $N$ values we simulate, a significant fraction of our configurations have sizeable order parameter values. That is, there is a significant presence of “pseudo-nematic domains.” The order parameter distribution this close to the transition converges very slowly to expected isotropic $s = 0$ delta function as $N$ increases, and neither the large-deviation nor CLT theories capture that slow convergence all that well. All we can say is that, even in this extreme case, our minimal large-deviation analysis continues to do noticeably better at predicting the large-fluctuation tails than the central-limit-theory does.

\begin{figure}
\centering
\includegraphics[width=0.60\textwidth, height=0.55\textheight]{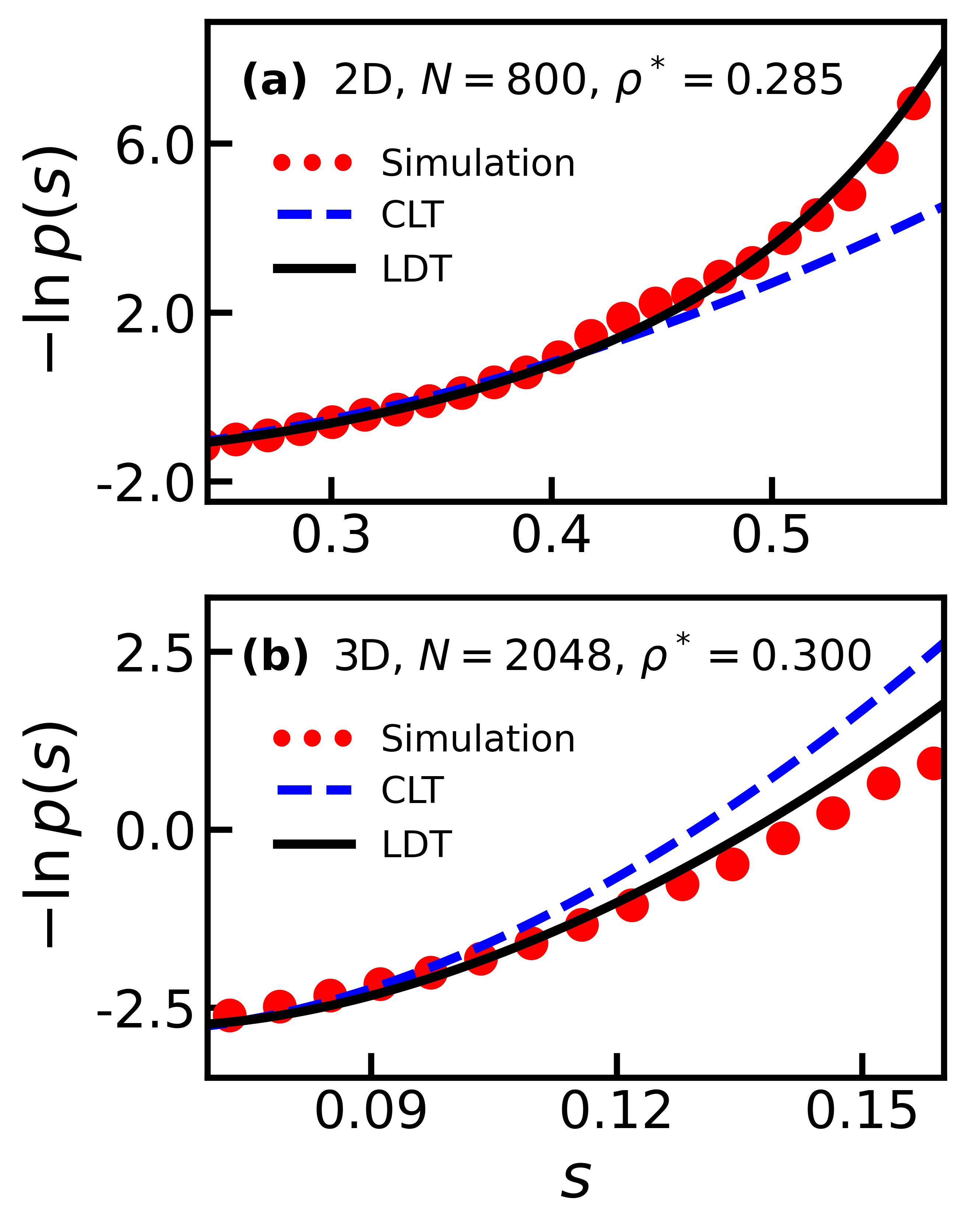}
\caption{Relative free energies associated with different levels of orientational order for isotropic Gay-Berne liquid crystals in the pre-transitional portions of their phase
diagrams. The upper and lower panels display the (negative logarithms of the) right-hand
tails of the (largest $N$) two- and three-dimensional probability densities shown in Figs. 5 and 6, respectively.}
\label{fig:7}
\end{figure}

\begin{figure}
\centering
\includegraphics[width=0.60\textwidth, height=0.55\textheight]{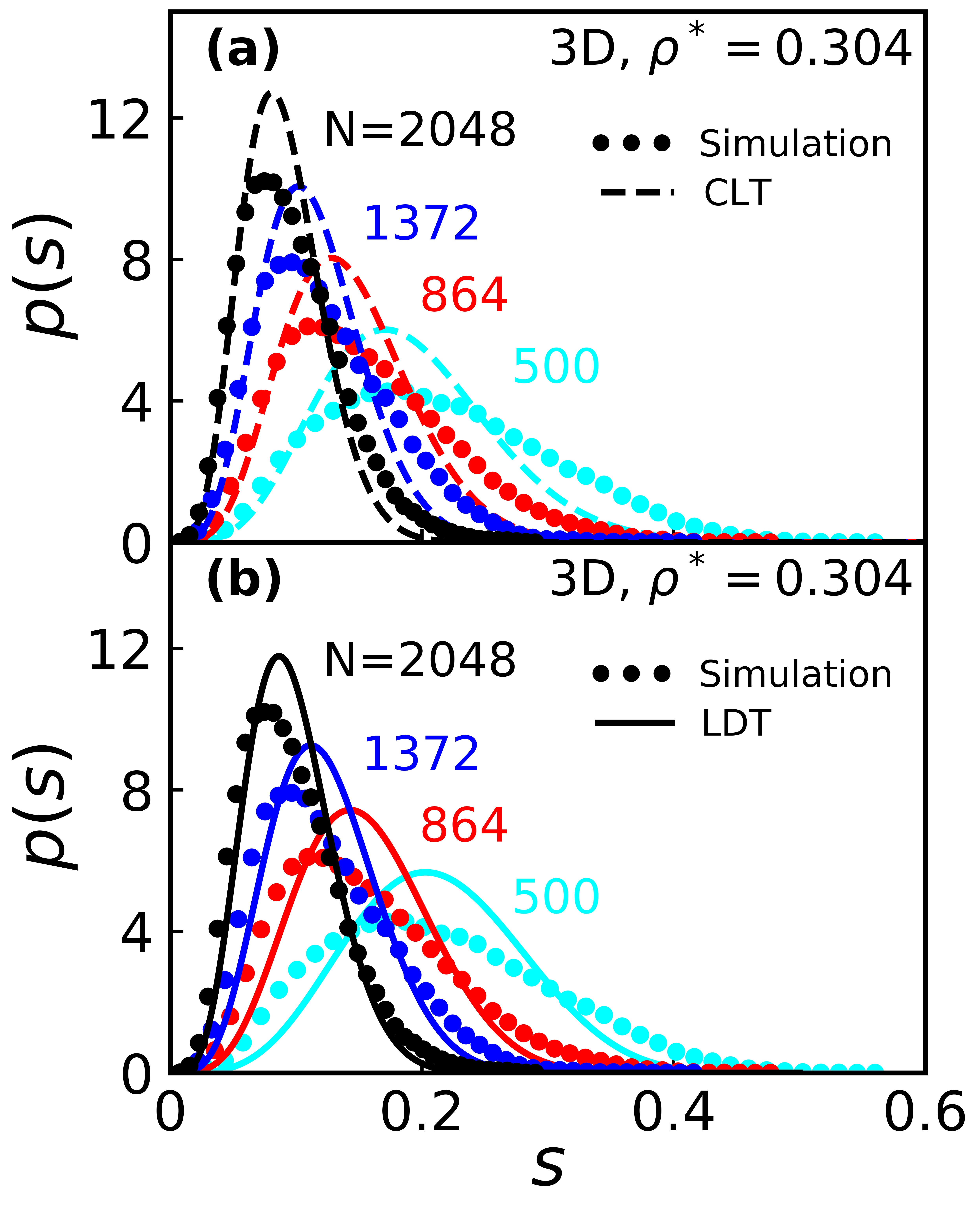}
\caption{Finite-size scaling results for the order-parameter probability distribution of an isotropic three-dimensional Gay-Berne liquid crystal in the extreme pre-transitional region of the phase diagram ($T^*= 1.00$, $\rho^*= 0.304$). All molecular dynamics results are averaged over $1.5 \times 10^7$ configurations. The upper and lower panels compare these simulation results with the central-limit theorem, and large-deviation-theory predictions, respectively. Notice, in particular, the differences in the right-hand tails of the two
theoretical predictions at larger $N$ values.}
\label{fig:8}
\end{figure}

\section{\label{sec:level1}CONCLUDING REMARKS}

The essential lesson of this work, at least for the classical orientational disorder problems being contemplated in this paper, is that the substantial simulation challenges posed by large fluctuations can be analytically redirected into simulation regimes that are much easier to access. The framework of large deviation theory reminds us that we can obtain the probability distribution of those fluctuations via a thermodynamic integration of the equation of state relating the target fluctuations to their conjugate thermodynamic field. What we point out here, though, is that the information needed to construct that equation of state is easiest to obtain by simulation just when it is most difficult to pursue analytically (when the system is strongly correlated but weakly fluctuating) and easiest to model analytically just when the system is hardest to simulate – because fluctuations are largest when the correlations are at their weakest. So an interpolation between these regimes can be effective in leveraging the computational information that we do have at our disposal well beyond its usual domain of reliability.

A useful perspective on why this approach might be particularly helpful comes from recognizing that by including nonlinear, largely single-molecule, behavior we are simply building in the correct \textit{boundary condition} that the entire distribution of fluctuations is subject to. But like boundary conditions in other mathematical contexts, such conditions can lead to effects felt throughout the full range of the problem. Large deviation theories are clearly better at getting at the tails of probability distributions than many other approaches are, but because we incorporating our boundary condition effects via an interpolation connecting us to the strongly-correlated part of the problem, we can aspire to predict \textit{both} the forms of those tails and any non-Gaussian aspects there might be to the overall character of the probability density.

Having said that, it is fair to ask whether we are doing no more than constructing a mean-field theory complete with all of the well-known limitations that such theories face when confronting near-critical thermodynamic conditions. The answer is yes and no. We believe that our expansion of the strong-field limit in Eqs. (4.15) and (4.32) is exact; we are aware of no reason why that limit should not be describable by a leading-order pole plus an analytical expansion about the perfectly \textit{ordered} state. This description gets better, not worse, as the fluctuations grow to their saturation values, so an interpolation from this limit is not automatically a mean-field theory. However, the
particular (Padé) choice for an interpolation scheme we introduced in this paper does impose a mean-field approximation; using a Padé is equivalent to postulating an analytical (Landau) expansion of fluctuations about the fully \textit{disordered} state. Such expansions are incapable of explaining non-classical critical behavior (much less explaining topological-defect-driven phenomena such as Kosterlitz-Thouless transitions) and thus have real limitations. We are fortunate that the approach to the avoided critical point in liquid crystalline systems does not seem to deviate from classical behavior until fairly close to the empirical critical point, but non-classical behavior eventually does set in \cite{stinson1970pretransitional, gramsbergen} – at which point our formalism becomes less useful. That may be at least part of what we see in Fig. 8.

It is worth pointing out that nothing in this development limits us to the specific $\sca{l} = 2$ orientational-order questions we dealt with in this paper. For example, the polymer end-to-end distance distribution problem \cite{Yamakawa, Rubinstein} also asks for the probability density of an extensive order-parameter vector, the end-to-end vector $\Delta \vect{R}$
\begin{equation*}
    \Delta \vect{R} = \sca{d} \displaystyle \sum_{j=1}^{N} \unitvector{\Omega}_j
\end{equation*}
whenever the polymer is composed of $N$ monomers of fixed length $d$, each with orientational unit vector $\unitvector{\Omega}_j$ . In its simplest guise, the random-flight (uncorrelated-monomer) limit, this $\sca{l} = 1$ problem is, of course, one of the most well-known problems in statistical physics,\cite{Dutka, lord} but the methods described in Secs. II and III allow us to extract the tail of the distribution from simulations without building in any restrictions on the level of intramolecular correlations the polymer might have.

In any case, now that we have the interpolated results for liquid-crystalline order, it is revealing to examine those results from the perspective afforded by random matrix theory. \cite{doerr, zhao} As discussed in our earlier work,\cite{zhao} finding the eigenvalue probability distributions of the $d \times d$ order-parameter matrices $\matr{Q}$ is a direct exercise in random matrix theory. One lesson from that analysis is that the only nonzero quantities invariant to similarity transformations are the $(\sca{d} - 1)$ quantities
\begin{equation*}
    Tr\left(\matr{Q}^2\right), Tr\left(\matr{Q}^3\right), ..., Tr\left(\matr{Q}^d\right)
\end{equation*}
or their equivalents.\footnote{Any permutation invariant combination of the $d$ eigenvalues, such as the determinant of $\matr{Q}$ is also an invariant quantity, but it is possible to show that any such invariant can be expressed as a function of the $Tr\left(\matr{Q}^n\right)$ quantities. For example, in two and three dimensions, $\text{det}(\matr{Q}) = - \frac{1}{2} Tr\left(\matr{Q}^2\right)$ and $\text{det}(\matr{Q}) = - \frac{1}{3} Tr\left(\matr{Q}^3\right)$, respectively.} Under isotropic conditions (when properties of the system must be invariant to similarity transformations such as rotations and reflections), the exact eigenvalue probability distribution can therefore depend on $\matr{Q}$ only through these quantities.

Indeed, in two dimensions we find our rate function, Eq. (4.18), depends just on $Tr\left(\matr{Q}^2\right)$ , as it has to. In particular, Eqs. (4.3a) and (4.5) imply that the scalar $\sca{Q}$ that appears in our rate function obeys $Tr\left(\matr{Q}^2\right) = 2Q^2$. In three dimensions, we also find that our rate function, Eq. (4.35), depends only on $Tr\left(\matr{Q}^2\right)$, because Eq. (4.36) tells us that the scalar $\sca{r}$ satisfies the relation $Tr\left(\matr{Q}^2\right) = \frac{3}{2} \sca{r}^2$ . However, our invariance analysis notes that the 3-$\sca{d}$ exact answer could also depend on $Tr\left(\matr{Q}^3\right)$ . We precluded that possibility in our development here by taking our field conjugate to $\matr{Q}$ to be a scalar $\sca{h}$. Was this assumption justified?

It is not difficult to see that the principal information in $Tr\left(\matr{Q}^3\right)$ missing from $Tr\left(\matr{Q}^2\right)$ arises from fluctuations transverse to the director: Without loss of generality, we can always write our traceless matrix $\matr{Q}$ in its eigenvector basis as
\begin{equation*}
    \begin{pmatrix}
    \ \sca{s} & 0 & 0 \\
    \ 0 & -\frac{\sca{s}}{2} + \delta & 0\\
    \ 0 & 0 & -\frac{\sca{s}}{2} - \delta\\
    \end{pmatrix}
\end{equation*}
so that the largest eigenvalue $\sca{s}$ measures the orientational fluctuations parallel to the director and $\delta$ measures the transverse fluctuations. If $\delta << \sca{s}$ , then
\begin{equation*}
    Tr\left(\matr{Q}^2\right)^3 = 6 \left( Tr\left(\matr{Q}^3\right) \right)^2
    \tag{6.1}\label{eq:6.1}
\end{equation*}
so under large-fluctuation conditions when the dominant fluctuations are nematic (as opposed to biaxial nematic), \cite{Gennes, Ilg2011} one would not expect any information to be missing from a theory that relies solely on $Tr\left(\matr{Q}^2\right)$.

Conversely, when the total extent of fluctuations is small, the central limit theorem applies, so, as we can see from both our earlier paper \cite{zhao} and the present work, the finite-size probability distribution of the order parameter matrix takes on the usual random-matrix-theory form
\begin{equation}
    \sca{P}(\matr{Q}) \sim \exp{\left(-\alpha Tr\left(\matr{Q}^2\right)\right)} \tag{6.2}\label{eq:6.2}
\end{equation}
in both two and three dimensions. The upshot of the numerical findings in this paper, then, are that the rapid growth of nematic-like domains, even in three dimensions, is largely captured by the evolution of $\sca{P}(\matr{Q})$ from the Gaussian form of Eq. (6.1) to a more general function of $Tr\left(\matr{Q}^2\right)$,
\begin{equation}
    \sca{P}(\matr{Q}) \sim f\left(Tr\left(\matr{Q}^2\right)\right) \tag{6.3}\label{eq:6.3}
\end{equation}
with limited contributions from the transverse fluctuations. The evolution of these domains, moreover, is determined to a surprising degree by how the simple nonlinearities inherent in individual molecular orientations act to shape the intermolecular correlations driving the isotropic-to-nematic transition.

\section*{SUPPLEMENTARY MATERIAL}

The supplementary material includes (1) the values of the $\chi$ and $\eta$ parameters used to generate our predicted probability distributions, and (2) an exploration of an alternative Padé approximant scheme.

\begin{acknowledgments}

We thank Brown University and the American Hellenic Educational Progressive Association for support provided to Eleftherios Mainas.

\end{acknowledgments}

\section*{Author Declarations}

\subsection*{Conflict of Interest}
The authors have no conflicts to disclose.
\subsection*{Authors Contributions}

\textbf{Eleftherios Mainas}: Conceptualization (equal); Methodology (equal); Software (lead); Visualization (lead); Writing (Original draft, review, and editing) (equal).

\textbf{Richard Stratt}: Conceptualization (equal); Methodology (equal); Writing (Original draft, review, and editing) (equal).

\section*{Data Availability Statement}

The data presented in the figures shown here are available from the authors on request.

\appendix

\section{Isotropic symmetry in two dimensions}
We can show that when our liquid is isotropic, the fluctuation matrix is proportional to the unit matrix times the susceptibility $\chi$
\begin{equation*}
    \langle \delta Q_{\mu} \delta Q_{\nu} \rangle = \langle Q_{\mu} Q_{\nu} \rangle - \langle Q_{\mu} \rangle \langle Q_{\nu} \rangle = \delta_{\mu \nu} \frac{\chi}{2N}, \quad \mu, \nu = x, y
\end{equation*}
To see this, note that an isotropic 2-$d$ system is invariant to rotating each molecule’s orientation $\theta_j$ by an identical angle $\alpha$, which implies that the only angle averages that survive involve angle differences. In particular, for all molecules $j, k$
\begin{equation*}
    \langle \cos{2\theta_j} \rangle = \langle \sin{2\theta_j} \rangle = \langle \cos{2(\theta_j + \theta_k)} \rangle = \langle \sin{2(\theta_j + \theta_k)} \rangle = 0
\end{equation*}
Thus
\begin{equation*}
    \langle Q_{\mu} \rangle = \frac{1}{N} \displaystyle \sum_{j=1}^{N} \left \langle \begin{pmatrix}
        \cos{2\theta_j} \\
        \sin{2\theta_j}
    \end{pmatrix}_{\mu} \right \rangle = 0
\end{equation*}
And since, independently of any isotropy,
\begin{equation*}
    \displaystyle \sum_{j,k} \sin{2(\theta_j - \theta_k)} = 0
\end{equation*}
we find from the definition of the 2-$d$ orientational susceptibility $\chi$ , Eq. (4.11)
\begin{equation*}
\begin{aligned}
    \langle Q_{\mu} Q_{\nu} \rangle &= \frac{1}{N^2} \displaystyle \sum_{j,k=1}^{N} \left \langle \begin{pmatrix}
        \cos{2\theta_j} \\
        \sin{2\theta_j}
    \end{pmatrix}_{\mu} 
    \begin{pmatrix}
        \cos{2\theta_k} \\
        \sin{2\theta_k}
    \end{pmatrix}_{\nu}
    \right \rangle \\
    &= \frac{1}{2N^2} \begin{cases}
        (\mu = \nu) \quad \displaystyle \sum_{j,k} \left( \left \langle \cos{2(\theta_j - \theta_k)} \pm \cos{2(\theta_j + \theta_k)} \right \rangle \right)\\
        (\mu \neq \nu) \quad \displaystyle \sum_{j,k} \left( \left \langle \sin{2(\theta_j - \theta_k)} + \sin{2(\theta_j + \theta_k)} \right \rangle \right)
    \end{cases} \\
    &= \frac{1}{2N^2} \begin{cases}
        (\mu = \nu) \quad \displaystyle \sum_{j,k} \left \langle \cos{2(\theta_j - \theta_k)} \right \rangle \\
        (\mu \neq \nu) \quad 0
    \end{cases} \\
    &= \delta_{\mu \nu} \frac{\chi}{2N}
\end{aligned}
\end{equation*}

\section{Tesseral harmonics and the implications of isotropic symmetry in three dimensions}

In 3-$d$, the Cartesian matrix elements of the $\matr{q}$ tensor, Eq. (4.1), for a molecule whose axis is pointed along the $\unitvector{\Omega}$ unit vector
\begin{equation*}
    \left( \matr{q} \right)_{\alpha} = \left(\frac{3}{2} \unitvector{\Omega} \unitvector{\Omega} - \frac{1}{2} \matr{1} \right)_{\alpha}
\end{equation*}
\begin{equation*}
    \alpha = 1(xx), \: \: 2(yy), \: \: 3(zz), \: \: 4(xy=yx), \: \: 5(xz=zx), \: \: 6(yz=zy)
\end{equation*}
\begin{equation*}
    q_1 + q_2 + q_3 = 0
\end{equation*}
can be expressed in terms of either the (complex) $l = 2$ spherical harmonics $Y_{2,m}(\unitvector{\Omega})$ or their real equivalent, the $l = 2$ tesseral harmonics, $r_{2,m}(\unitvector{\Omega}), (m = -2, ...,2)$. Dropping the $l$ part of the subscript for notational simplicity,
\begin{equation*}
\begin{aligned}
    &r_2 = \frac{1}{\sqrt{2}}(Y_2 + Y_{-2}) = \sqrt{\frac{5}{12\pi}} (q_1 - q_2) \\\\
    &r_1 = \frac{1}{\sqrt{2}}(Y_{-1} - Y_1) = \sqrt{\frac{5}{3\pi}} q_5 \\\\
    &r_0 = Y_0 = \sqrt{\frac{5}{4\pi}} q_3 = - \sqrt{\frac{5}{4\pi}}(q_1 + q_2) \\\\
    &r_{-1} = \frac{i}{\sqrt{2}}(Y_1 + Y_{-1}) = \sqrt{\frac{5}{3\pi}} q_6 \\\\
    &r_{-2} = \frac{i}{\sqrt{2}}(Y_{-2} - Y_2) = \sqrt{\frac{5}{3\pi}} q_4
\end{aligned}
\end{equation*}
The same relationships hold between the corresponding $N$-body quantities $R_m$ and $Q_{\alpha}$ , Eqs. (4.22) and (4.21). So, from Eq. (4.20) we can show that Eq. (4.24) holds
\begin{equation*}
    Tr\left(\matr{Q}^2\right) = Q_1^2 + Q_2^2 + Q_3^2 + 2(Q_4^2 + Q_5^2 + Q_6^2) = \frac{6\pi}{5} \displaystyle \sum_{m=-2}^{2} R_m^2
\end{equation*}

Tesseral harmonics obey the same kinds of spherical harmonic addition theorems that spherical harmonics themselves do. In particular, our $l = 2$ functions obey
\begin{equation*}
    \displaystyle \sum_{m=-2}^{2} r_m(\unitvector{\Omega}_j) r_{m}(\unitvector{\Omega}_k) = \displaystyle \sum_{m=-2}^{2} Y_m^*(\unitvector{\Omega}_j) Y_m^*(\unitvector{\Omega}_k) = \frac{5}{4\pi} P_2(\unitvector{\Omega}_j \cdot \unitvector{\Omega}_k)
\end{equation*}
Moreover, when our system is isotropic, tesseral harmonics also average much the same way that spherical harmonics do. For example, isotropy of our liquids implies that every observable must invariant to rotating each molecule’s azimuthal angle $\phi_j$ by an identical angle $\alpha$. Since each $r_m(\Omega) = r_m(\theta, \phi) \sim \cos{m\phi}$ or $\sin{m\phi}$ , that means that pair averages involving different $m$ values must vanish
\begin{equation*}
    \left \langle r_m(\unitvector{\Omega}_j) r_{m'}(\unitvector{\Omega}_k) \right \rangle_{m \neq m'} = 0
\end{equation*}

These last two properties can be used to establish Eq. (4.26) since they imply a connection with the 3-$d$ orientational susceptibility $\chi$, Eq. (4.27)
\begin{equation*}
    \frac{1}{N} \displaystyle \sum_{j,k}^{N} \displaystyle \sum_{m,m'=-2}^{2} \left \langle r_m(\unitvector{\Omega}_j) r_{m'}(\unitvector{\Omega}_k) \right \rangle = \frac{5}{4\pi} \left \langle \frac{1}{N} \displaystyle \sum_{j,k}^{N} P_2(\unitvector{\Omega}_j \cdot \unitvector{\Omega}_k) \right \rangle = \frac{5}{4\pi} \chi
\end{equation*}

\section{The relationship between our work and the Landau free energy calculations of Ilg and co-workers}

The presence of some elements of overlap between the work described in this paper and that of Ilg and coworkers \cite{Ilg2011, Ilg2012, Gupta2013, Luo2014} makes it worthwhile to spell out a few of the similarities and differences between our respective approaches. Part of those authors’
efforts, for example, is devoted to an accurate representation of the order-parameter dependence of the entropy of ideal (non-interacting) two \cite{Luo2014} and three \cite{Ilg2011} -dimensional
liquid-crystal systems. As we do, they explicitly replace the order-parameter information with that provided by a Laplace-transform conjugate field, explicitly consider both the weakly and strongly ordered limits, and use the conjugate field to guide an extrapolation between the two limits. They further make a point, as we do, of including the logarithmic singularity in the entropy in the maximally ordered limit.

However, there are both conceptual and practical differences between our formulations: (1) The large deviation perspective presented in this paper allows us to address finite-size scaling issues explicitly, whereas simply assuming a Landau free energy form does not. (2) When treating fully interacting systems we are not limited to free energies of the traditional mean-field form – an ideal entropy plus an average energy. Interactions make themselves felt in the entropy, as well as in the energy, because those interactions raise the values of the susceptibilities $\chi$. Our development allows for this interaction-induced renormalization of the entropy (see, e.g., Eq. (3.2)). (3) The interpolation we carry out between the weakly and strongly ordered limits is not an arbitrary numerical fitting; it is completely defined by our requirement for the simplest possible mathematical form respecting the symmetry of the system and the known exact behaviors of the (order-parameter) equation of state at those two limits.

The fact that we use an equation-of-state interpolation and that Ilg and coworkers
do a free energy interpolation is not itself an important consideration; the two are
formally equivalent. But the equation-of-state version makes it somewhat easier to have the interpolation respect the proper symmetry. We note, for example the two-dimensional Landau free energy study \cite{Luo2014} writes their interpolated ideal entropy as an impressive looking 7-term function of order parameter $s$,
\begin{equation}
    S_0(s)/(Nk_B) = \frac{1}{2}\ln{(1-s)} + \frac{1}{2}s - \frac{3}{4}s^2 + \frac{1}{6}s^3 - \frac{1}{8}s^4 + \frac{1}{10}s^5 - \frac{1}{18}s^6
    \tag{C.1}\label{eq:C.1}
\end{equation}
whereas simply having the equation of state taking into account that s must be an odd function of the system’s conjugate field, Eq. (4.16), immediately leads to an entropy that correctly builds in the invariance of the entropy to the sign of $s$, Eq. (4.18a):
\begin{equation}
    S_0(s)/(Nk_B) = \frac{1}{2}\ln{(1-s^2)} - \frac{1}{2}s^2 + \frac{1}{8}s^4
    \tag{C.2}\label{eq:C.2}
\end{equation}
The reader can easily verify that Eq. (C.2) exactly captures all but the 6th order contribution to Eq. (C.1). All of the odd order polynomial terms in Eq. (C.1) serve just to cancel out the first 3 odd order terms coming from the expansion of the logarithm.

Our respective treatments of the analogous three-dimensional problem \cite{Ilg2011} raise
some other issues worth noting. Under the assumption that only ordering possible is of the nematic variety, we again end up with very similar expressions for the ideal entropies inherent in our two theories (aside from some cosmetic notational differences involving rescaled definitions of the basic order parameter tensor, and mathematically equivalent choices between writing Dawson’s function and the imaginary error function). The limiting ideal-equation-of-state expressions in this work, Eq. (4.32) (with the ideal value $\eta$), for example, are identical to Ilg et al.’s disordered $(s \rightarrow 0)$ and nematically ordered $(s \rightarrow 1)$ limits \cite{Ilg2011} in their Eq. (B.11).

More generally, our Eq. (6.1) connecting cubic and quadratic dependencies on the order-parameter tensor is exact provided the system has no significant non-uniaxial fluctuations. However, Ilg and coworkers \cite{Ilg2011} correctly point out that, even within this restriction, a liquid-crystalline system could also feature discotic order $(s \rightarrow -\frac{1}{2})$, a possibility this paper did not take into account. The phase diagram of the particular Gay-
Berne model we tested our ideas on does not appear to support an equilibrium discotic phase under our simulated thermodynamic conditions, \cite{demiguel} but, even so, it is possible that we could improve our treatment of improbable fluctuations by at least allowing for discotic fluctuations. In our context, that would simply mean including the discotic pole
in our equation-of-state Padé, which would give us the additional logarithmic singularity in the ideal entropy included by Ilg and coworkers.\cite{Ilg2011}

As a final point, we should mention that Ilg and coworkers also show how one can nicely incorporate the possibility of biaxial ordering (the system simultaneously selecting two different ordering axes) by generalizing the conjugate field to a second-rank tensor.\cite{Ilg2011} Generalizing the vector version of large deviation development, Eqs. (2.6)-(2.8), in this way would straightforwardly parallel published work, \cite{zhao, Ilg2011} but it remains to be established how well the resulting equation of state would continue to be represented
by a Padé approximant.

\bibliography{mybib}

\end{document}